\documentclass[a4paper]{cas-sc}

\usepackage[numbers,sort&compress]{natbib}
\usepackage[version=4]{mhchem}
\usepackage{siunitx}
\usepackage{bm}
\usepackage{float}
\usepackage{layouts} 

\usepackage[utf8]{inputenc}
\usepackage{hyperref}
\DeclareSIUnit{\angstrom}{\text{\AA}}
\newcommand{\openone}{\leavevmode\hbox{\small1\kern-3.8pt\normalsize1}}

\AddToHook{env/Abstract/before}{%
    \par
    \vspace{0.4em}

    \begingroup
        \parindent=0pt

        \noindent
        G\,R\,A\,P\,H\,I\,C\,A\,L\quad
        A\,B\,S\,T\,R\,A\,C\,T
        \par

        \vspace{-4pt}

        \noindent
        \rule{\textwidth}{.2pt}
        \par
    \endgroup

    \vspace{0.5em}

    \noindent
    \includegraphics[
        width=\textwidth,
        height=0.3\textheight,
        keepaspectratio
    ]{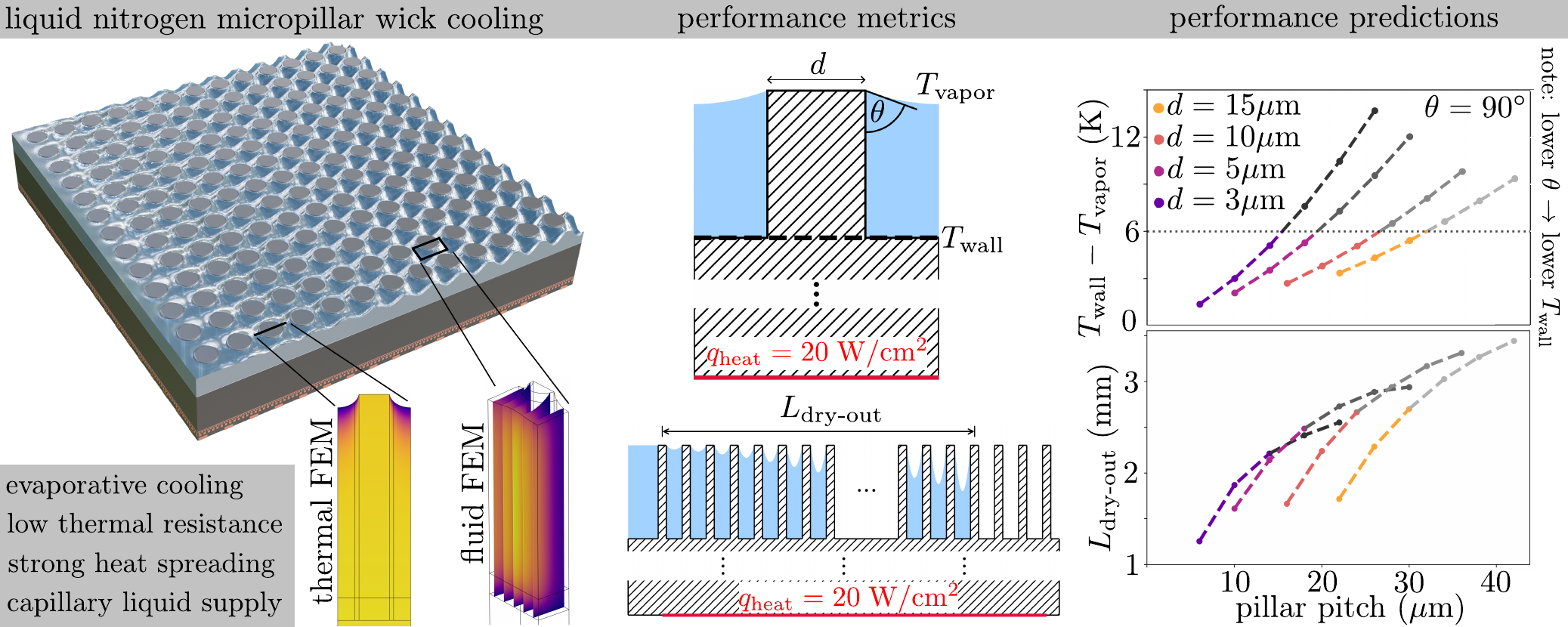}

    \par
    \vspace{0.8em}
}

\def\tsc#1{\csdef{#1}{\textsc{\lowercase{#1}}\xspace}}
\tsc{WGM}
\tsc{QE}

\begin{document}
\let\WriteBookmarks\relax
\def\floatpagepagefraction{1}
\def\textpagefraction{.001}

\shorttitle{}    

\shortauthors{}  

\title[
  mode = title,
  size = \Large
]{Liquid-Nitrogen Micropillar-Wick Cooling for Cryogenic Electronics: \newline A Numerical Study of Thermal Performance and Capillary Dry-Out Limits}  



%


\author[1, 2]{Felix Mende}[orcid=0009-0002-5840-0181]

\ead{felix.mende@ipms.fraunhofer.de}
\cormark[1]

\author[1]{Marcus Wislicenus}[orcid=0009-0005-2678-8144]

\author[3]{Dianping Jiang}[orcid=0000-0002-5353-0617]


\author[3]{Munehiro Tada}[orcid=0000-0002-1015-2222]

\author[2,4]{Lukas M. Eng}[orcid=0000-0002-2484-4158]

\affiliation[1]{
    organization={Fraunhofer Institute for Photonic Microsystems IPMS},
    addressline={Center Nanoelectronic Technologies (CNT)},
    postcode={01109},
    city={Dresden},
    country={Germany}
}

\affiliation[2]{
    organization={Institute of Applied Physics, TU Dresden},
    addressline={N{\"o}thnitzer Strasse 61},
    postcode={01187},
    city={Dresden},
    country={Germany}
}

\affiliation[3]{
    organization={Keio University},
    city={Yokohama},
    country={Japan}
}

\affiliation[4]{
    organization={ctd.qmat: Dresden-W{\"u}rzburg Cluster of Excellence---EXC 2147, TU Dresden},
    postcode={01062},
    city={Dresden},
    country={Germany}
}



\begin{abstract}
Cryogenic computing technologies are maturing rapidly, but heat removal remains a key challenge when increasing the device density and operating power. Two-phase evaporative cooling is a promising approach to solve this issue, because it can dissipate high heat fluxes while maintaining small temperature rises. Here, we numerically investigate liquid-\ce{N2}-filled silicon micropillar wicks as a capillary-fed thin-film evaporation concept for cryogenic electronics. The model combines Young-Laplace meniscus calculations, Hertz-Knudsen-Schrage evaporation, unit-cell heat-transfer and liquid-flow simulations, and an array-level thermal and capillary-flow model. For a representative geometry with a pillar diameter of $\qty{10}{\micro\meter}$, pitch of $\qty{24}{\micro\meter}$, and pillar height of $\qty{75}{\micro\meter}$ at an applied heat flux of $\qty{20}{\watt\per\centi\meter\squared}$, the predicted chip-temperature rise is approximately $\qty{2.7}{\kelvin}$. This is substantially below the estimated temperature rises for representative conduction cooling through an indium-interlayered copper heat sink and direct liquid-\ce{N2} immersion. The corresponding predicted single-fed dry-out length is approximately $\qty{2.7}{\milli\meter}$, equivalent to an ideal double-fed coolable width of approximately $\qty{5.3}{\milli\meter}$. The results indicate that local thermal performance is favorable, whereas lateral scalability is primarily constrained by capillary dry-out. An approximate capillary-viscous scaling relation provides a compact framework for comparing dry-out limits across working fluids and related wick geometries and for identifying strategies to extend capillary-fed transport.
\end{abstract}


\maketitle


\AddToHook{env/Abstract/before}{%
    \par
    \vspace{0.4em}

    \begingroup
        \parindent=0pt

        \noindent
        G\,R\,A\,P\,H\,I\,C\,A\,L\quad
        A\,B\,S\,T\,R\,A\,C\,T
        \par

        \vspace{-4pt}

        \noindent
        \rule{\textwidth}{.2pt}
        \par
    \endgroup

    \vspace{0.5em}

    \noindent
    \includegraphics[
        width=\textwidth,
        height=0.3\textheight,
        keepaspectratio
    ]{figs/LN2_Paper_GraphicalAbstract.pdf}

    \par
    \vspace{0.8em}
}

\setlength{\fboxsep}{8pt}   
\setlength{\fboxrule}{0.5pt} 

\noindent
\fbox{%
\begin{minipage}{\dimexpr\textwidth-2\fboxsep-2\fboxrule\relax}
\section*{Nomenclature}

\small
\begin{minipage}[t]{0.485\linewidth}
\vspace{0pt}
\begin{tabular*}{\linewidth}{@{\extracolsep{\fill}}lll@{}}
\toprule
Symbol & Definition & Unit \\
\midrule
\multicolumn{3}{@{}l}{\textit{Geometry and array variables}} \\
$d$ & pillar diameter & \si{\meter} \\
$l$ & pillar pitch; unit-cell length & \si{\meter} \\
$h$ & pillar height & \si{\meter} \\
$N$ & number of unit cells in $x$-dir. & -- \\
$n$ & unit-cell index & -- \\
$x,y,z$ & Cartesian coordinates & -- \\
$L$ & total wick length & \si{\meter} \\
\midrule
\multicolumn{3}{@{}l}{\textit{Pressure and capillarity}} \\
$p_{\mathrm l}$ & liquid pressure & \si{\pascal} \\
$p_{\mathrm v}$ & vapor pressure & \si{\pascal} \\
$p_{\mathrm{l,edge},n}$ & liquid pressure at cell $n$ edge & \si{\pascal} \\
$p_{\mathrm l,n}$ & liquid pressure at cell $n$ center & \si{\pascal} \\
$\Delta p_{\mathrm{cell}}$ & pressure drop across unit cell & \si{\pascal} \\
$P$ & Laplace pressure, $p_{\mathrm v}-p_{\mathrm l}$ & \si{\pascal} \\
$P_{\mathrm{crit}}$ & critical Laplace pressure & \si{\pascal} \\
$P_{\mathrm{edge},n}$ & Laplace pressure at cell edge $n$  & \si{\pascal} \\
$P_{n}$ & Laplace pressure at cell center $n$  & \si{\pascal} \\
$\sigma_{\mathrm{lv}}$ & liquid--vapor surface tension & \si{\newton\per\meter} \\
$\theta$ & contact angle & \si{\degree} \\
$\theta_{\mathrm{r}}$ & receding contact angle & \si{\degree} \\
\midrule
\multicolumn{3}{@{}l}{\textit{Flow variables and material properties}} \\
$\bm v$ & liquid velocity field & \si{\meter\per\second} \\
$\rho_{\mathrm l}$ & liquid mass density & \si{\kilogram\per\meter\cubed} \\
$\rho_{\mathrm v}$ & vapor mass density & \si{\kilogram\per\meter\cubed} \\
$\mu_{\mathrm l}$ & liquid dynamic viscosity & \si{\pascal\second} \\
$\underline K$ & viscous stress tensor & \si{\pascal} \\
$\underline I$ & identity tensor & -- \\
$\hat{\bm n}$ & outward unit-normal vector & -- \\
$\dot m_{\mathrm{liq},n}$ & mass flow rate through cell $n$ & \si{\kilogram\per\second} \\
$R_{\mathrm{cell}}$ & unit-cell mass-flow resistance & \si{\pascal\second\per\kilogram} \\
\bottomrule
\end{tabular*}
\end{minipage}%
\hfill
\begin{minipage}[t]{0.485\linewidth}
\vspace{0pt}
\begin{tabular*}{\linewidth}{@{\extracolsep{\fill}}lll@{}}
\toprule
Symbol & Definition & Unit \\
\midrule
\multicolumn{3}{@{}l}{\textit{Mass transfer and Schrage model}} \\
$j_{\mathrm{lv}}$ & net interfacial mass flux & \si{\kilogram\per\meter\squared\per\second} \\
$\dot m_{\mathrm{evap},n}$ & evaporated mass flow from cell $n$ & \si{\kilogram\per\second} \\
$\dot m_{\mathrm{evap,total}}$ & total evaporated mass flow & \si{\kilogram\per\second} \\
$h_{\mathrm{lv}}$ & specific enthalpy of vaporization & \si{\joule\per\kilogram} \\
$\bar v_{\mathrm n}$ & mean normal vapor drift velocity & \si{\meter\per\second} \\
$\Gamma$ & velocity-shift correction factor & -- \\
$\sigma_{\mathrm{ac}}$ & mass accommodation coefficient & -- \\
$\sigma_{\mathrm e}$ & evaporation coefficient & -- \\
$\sigma_{\mathrm c}$ & condensation coefficient & -- \\
$p_{\mathrm{sat}}(T_{\mathrm l})$ & saturation vapor pressure & \si{\pascal} \\
$k_{\mathrm{B}}$ & Boltzmann constant & \si{\joule\per\kelvin} \\
\midrule
\multicolumn{3}{@{}l}{\textit{Thermal variables}} \\
$T$ & temperature & \si{\kelvin} \\
$T_{\mathrm l}$ & liquid-interface temperature & \si{\kelvin} \\
$T_{\mathrm v}$ & far-field vapor temperature & \si{\kelvin} \\
$\bar T_{\mathrm{wall}}$ & mean wall temperature & \si{\kelvin} \\
$\Delta T$ & wall superheat  $\bar T_{\mathrm{wall}} - T_{\mathrm{v}}$ & \si{\kelvin} \\
$\bm q$ & local heat-flux vector & \si{\watt\per\meter\squared} \\
$q_{\mathrm{heat}}$ & imposed heat flux at substrate & \si{\watt\per\meter\squared} \\
$q_{\mathrm{CHF}}$ & critical heat flux & \si{\watt\per\meter\squared} \\
$q_{\mathrm{HKS}}$ & evaporative heat flux (HKS model) & \si{\watt\per\meter\squared} \\
$q_{\mathrm{LFOM}}$ &  liquid figure of merit & \si{\watt\per\meter\squared}  \\
$\kappa$ & thermal conductivity & \si{\watt\per\meter\per\kelvin} \\
$R_{\mathrm{eff}}$ & areal effective thermal resistance & \si{\kelvin\meter\squared\per\watt} \\
\midrule
\multicolumn{3}{@{}l}{\textit{Dimensionless groups and constants}} \\
$C_{\mathrm K}$ & Kutateladze coefficient & -- \\
$\mathrm{M}$ & Mach number based on $\bar v_{\mathrm n}$ & -- \\
$\mathrm{Ma}$ & Marangoni number & -- \\
$g$ & gravitational acceleration & \si{\meter\per\second\squared} \\
$\mathrm{Bo}$ & Bond number & -- \\
\bottomrule
\end{tabular*}
\end{minipage}%
\end{minipage}%
}
\vspace{1.5em}


\section{Introduction}
\label{sec}

Beyond its established role in quantum-control electronics \cite{ Xue.2021, Pauka.2021}, cryogenic operation of silicon electronics is increasingly considered for conventional high-performance computing \cite{Zota.2024,Lu.2026}. Operation near the normal boiling point of liquid \ce{N2} $(\approx\qty{77}{\kelvin})$ can improve CMOS power-performance characteristics through steeper subthreshold swing, enhanced carrier transport, reduced leakage, and lower interconnect resistance \cite{Lu.2026,Chiang.2020}. For a $\qty{10}{\nano\meter}$ FinFET technology, Chiang et al. combined low-temperature device measurements with circuit analysis and reported logic power of approximately $0.27\times$ the room-temperature value at constant speed, or a speed increase of approximately $\qty{49}{\percent}$ at constant power \cite{Chiang.2020}. Lee et al. likewise projected reductions in data-center power cost of up to $\qty{13.8}{\percent}$ for cryogenic-memory configurations \cite{Lee.2019}. 

Realizing these benefits requires control of the device temperature under load. Finite thermal resistance causes the active silicon to operate above the cryogenic reservoir temperature. In particular, cryogenic CMOS experiments have reported self-heating and thermal-coupling-induced temperature rises of several tens of Kelvin at a \qty{77}{\kelvin} ambient \cite{Bergamaschi.2024,Hart.2021,Vanbrabant.2025,Parihar.2026}. Such temperature excursions can substantially degrade cryogenic device characteristics \cite{Vanbrabant.2025}. Low thermal resistance between the active silicon and the cryogenic reservoir is therefore required both to preserve low-temperature device performance and to sustain high heat fluxes within a prescribed device-temperature limit.

Conventional cryogenic cooling approaches include conduction through a mechanically coupled heat sink and direct immersion in liquid \ce{N2}. Conduction cooling introduces additional solid-solid thermal interfaces between the device and the heat sink. Direct immersion avoids these additional interfaces but is ultimately limited by the critical heat flux (CHF). Capillary-fed micropillar wicks provide an alternative based on thin-film evaporation. Water-filled micropillar wicks have demonstrated low thermal resistances and high evaporative heat fluxes \cite{Adera.2016,Zhu.2016,Vaartstra.2019}. Compared to water, liquid \ce{N2} has a lower viscosity, which reduces viscous pressure losses. However, its lower surface tension reduces the available capillary pressure, while its lower latent heat requires a larger replenishment mass flow to remove the same heat load.

Liquid-\ce{N2} evaporation from stainless-steel pillars with square cross sections has recently been investigated experimentally for a pillar side length of $\qty{400}{\micro\meter}$ and a pitch of $\qty{900}{\micro\meter}$ at an elevated pressure of $\qty{1.38}{\mega\pascal}$ \cite{Hasan.2025}. Here, we instead consider dense arrays of cylindrical silicon micropillars with diameters $\leq\qty{15}{\micro\meter}$ and characteristic spacings below $\qty{50}{\micro\meter}$ at $\qty{101.3}{\kilo\pascal}$.
In this regime, capillary pressure, viscous resistance, meniscus geometry, and thin-film evaporation differ substantially from those of larger pillar structures. To date, a coupled thermal and capillary-flow description of liquid-\ce{N2} evaporation in this dense micrometer-scale regime remains insufficiently established.

We numerically investigate liquid-\ce{N2}-filled silicon micropillar wicks for electronics operated near \qty{77}{\kelvin}. We first compare their thermal response with representative conduction-cooled and direct-immersion configurations. A multiscale model then couples unit-cell calculations of meniscus shape, Hertz-Knudsen-Schrage evaporation, heat transfer, and mass flow resistance to an array-level model of temperature, and liquid flow. Finally, we derive an approximate capillary-viscous scaling relation for comparing dry-out limits across related wick geometries and working fluids.

\section{Comparison to conventional cryogenic cooling}
\label{sec:cooling_benchmarks}

\begin{figure}[pos=h]
\centering
\includegraphics[width=1\linewidth]{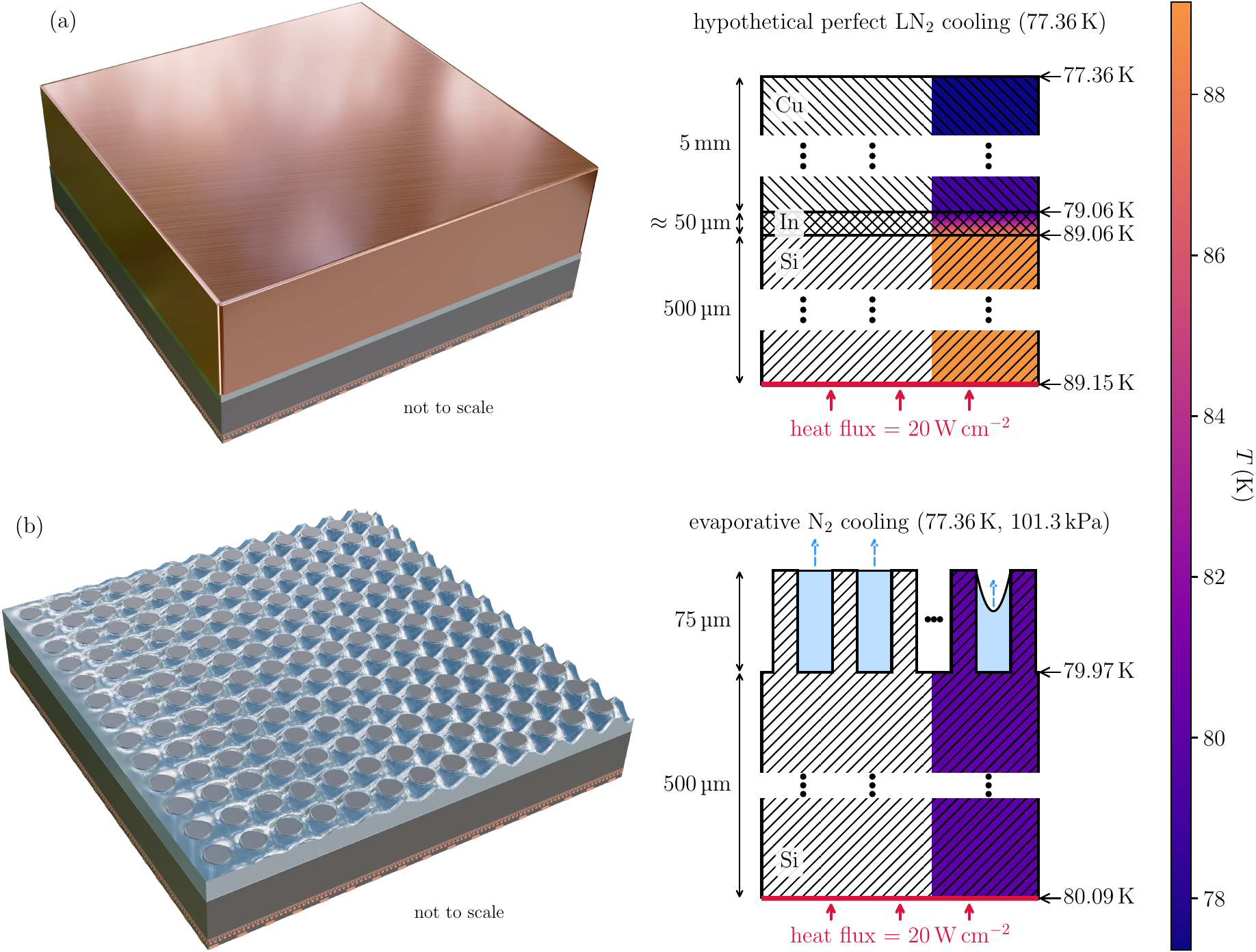}
\caption{Representative temperature profiles for two cryogenic cooling configurations under saturated liquid-\ce{N2} conditions ($\qty{77.36}{\kelvin}$, $\qty{101.3}{\kilo\pascal}$). Left: schematic of the physical structure. Right: corresponding geometry and temperature profile. (a) Conduction cooling through a $\qty{50}{\micro\meter}$ In interlayer between Si and an ideally heat-sunk Cu substrate. (b) Capillary-fed thin-film evaporation from a silicon micropillar wick. Profiles correspond to $q_{\mathrm{heat}}=\qty{20}{\watt\per\centi\meter\squared}$.}
\label{fig:TemperatureProfileComparison}
\end{figure}

To quantify the thermal-management considerations introduced above, we compare three representative cooling configurations: (i) conduction cooling through an indium-interlayered heat sink, (ii) flat-surface immersion in saturated liquid \ce{N2}, and (iii) capillary-fed thin-film evaporation from a micropillar wick. The comparison is intended to establish characteristic thermal-resistance and heat-flux levels rather than optimized system-level designs.

For the conduction-cooled reference, the \ce{Si} chip is mechanically coupled to an ideally heat-sunk \ce{Cu} substrate through a pressed In foil [see Fig.~\ref{fig:TemperatureProfileComparison}(a)]. Such a joint provides a representative high-conductance benchmark for mechanically coupled cryogenic interfaces \cite{Ekin.2006}. Still higher conductances can be achieved with permanently soldered joints \cite{Radebaugh.1977,Ekin.2006}. 
Using an effective \ce{Si}-\ce{In}-\ce{Cu} thermal interface conductance of \qty{20}{\kilo\watt\per\meter\squared\per\kelvin} \cite{Mochizuki.2007} for a pressed indium contact, the interface dominates the temperature rise because the thermal conductivities of both \ce{Si} and \ce{Cu} are comparatively high near $\qty{77}{\kelvin}$. At an applied heat flux of \qty{20}{\watt\per\centi\meter\squared}, the resulting temperature rise at the representative heating plane is approximately \qty{12}{\kelvin}.

Direct immersion removes this solid-solid interface, but introduces a different constraint. For an flat surface cooled by saturated liquid \ce{N2} under normal boiling conditions, the cryogenic critical-heat-flux (CHF) correlation of Patel et al.~\cite{Patel.2022} gives
\begin{equation}
\label{eq:flat_surface_LN2_CHF}
q_{\mathrm{CHF}}
=
\qty{19.3 \pm 4.7}{\watt\per\centi\meter\squared} \;,
\end{equation}
where the uncertainty follows from the relative RMS deviation reported for the underlying cryogenic pool-boiling database (Appendix~\ref{sec:CHF_details}). Thus, the representative heat flux of \qty{20}{\watt\per\centi\meter\squared} already approaches the expected CHF of a flat liquid-\ce{N2}-cooled surface. This limitation is consistent with the historical experience of liquid-\ce{N2}-cooled CMOS systems, in which CHF constituted an explicit thermal-design constraint \cite{Carlson.1989}.

A finite wall superheat is also required below CHF. Classical liquid-\ce{N2} pool-boiling measurements compiled by Richards et al.~\cite{Richards.1961} and summarized by Ekin~\cite{Ekin.2006} can be represented approximately by
\begin{equation}
\label{eq:LN2_nucleate_boiling}
q_{\mathrm{heat}}
=
C_{\mathrm{nuc}}\Delta T^{2.5} \;,
\end{equation}
where $q_{\mathrm{heat}}$ is the imposed heat flux, $C_{\mathrm{nuc}}$ depends strongly on surface condition and orientation and typically lies between \qty{0.01}{\watt\per\centi\meter\squared\per\kelvin^{2.5}} and \qty{0.1}{\watt\per\centi\meter\squared\per\kelvin^{2.5}}. Even adopting the upper end of this range, \qty{20}{\watt\per\centi\meter\squared} corresponds to a wall superheat of approximately \qty{8.3}{\kelvin}. Flat-surface immersion therefore removes the solid thermal interface but does not necessarily maintain the heated surface close to the liquid-\ce{N2} saturation temperature.

The micropillar-wick configuration instead transfers heat through a thin liquid-\ce{N2} film to a pinned liquid-vapor interface [Fig.~\ref{fig:TemperatureProfileComparison}(b)]. Liquid removed by evaporation is replenished by capillary flow from an adjacent reservoir. For the representative geometry considered here, the calculated temperature rise at \qty{20}{\watt\per\centi\meter\squared} is about \qty{2.7}{\kelvin}. This temperature rise is substantially smaller than those predicted for the conduction and flat-surface immersion benchmarks. The predicted temperature rise depends sensitively on the accommodation coefficient used in the Schrage model (see Sec.~\ref{sec:thermal_simulation_unit_cell}) its estimation is discussed in Appendix~\ref{sec:accomodation_coefficient}.

\section{Cell-level model}

\begin{figure}[pos=htbp]
    \centering
    \includegraphics
    {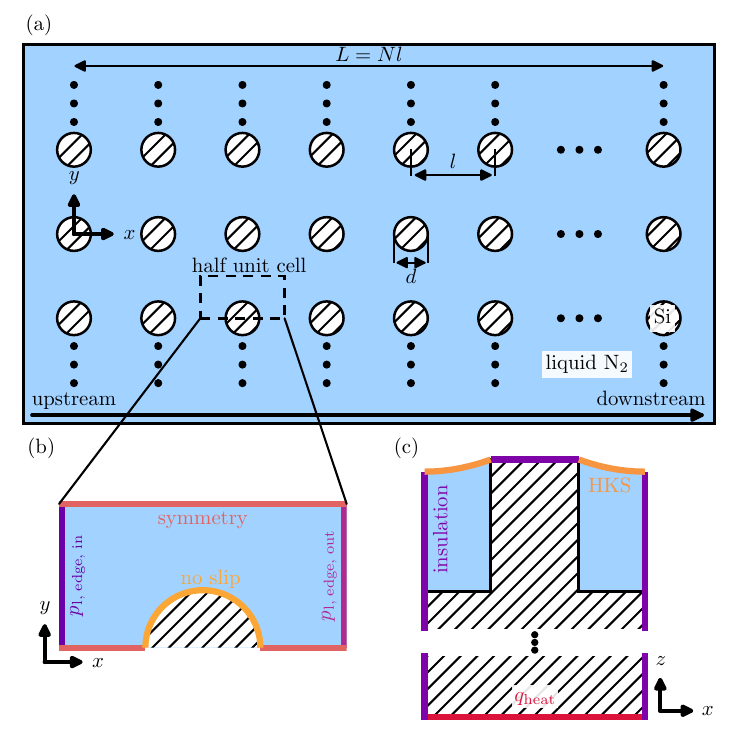}
    \caption{Unit-cell geometry and boundary conditions. (a) Square micropillar-array geometry and notation. (b) Half-cell liquid-flow domain with transverse symmetry, no-slip at the silicon surface, and the prescribed streamwise pressure drop. (c) Thermal domain with uniform heat flux $q_{\mathrm{heat}}$ at the substrate bottom, insulated lateral and upper silicon boundaries, and Hertz-Knudsen-Schrage evaporation at the liquid-vapor interface.}
    \label{fig:Geometry_BC_sketch}
\end{figure}

To quantify the thermal and capillary performance of the micropillar-wick configuration, we compute steady-state unit-cell solutions for square silicon micropillar-wick arrays. The unit-cell calculations provide the local liquid-vapor interface shape, thermal response, evaporation rate, and liquid-flow resistance for prescribed geometry, applied heat flux, and Laplace pressure.

To reduce computational cost, the coupled interfacial, thermal, evaporative, and liquid-flow problem is solved sequentially using a one-way-coupled workflow. First, the liquid-vapor interface shape is computed for a prescribed Laplace pressure. The temperature field is then solved in the resulting geometry. Finally, the temperature-dependent liquid properties are evaluated from this thermal solution and used in the liquid-flow simulation. This approach captures the influence of non-isothermal material properties while avoiding a fully coupled multiphysics calculation. The implications of this approximation are discussed below and in the appendices.

\subsection{Liquid-vapor interface shape and critical pressure}
\label{sec:Young_Laplace_equation}

Heated micropillar wicks exhibit curved liquid-vapor interfaces whose apparent curvature increases with distance from the liquid reservoir and with increasing applied heat flux \cite{Antao.2016}. This behavior is described by the Young-Laplace equation,
\begin{equation}
\label{eq:Young_Laplace}
\nabla_{\mathrm{s}} \cdot \hat{\bm n}
=
-\frac{P}{\sigma_{\mathrm{lv}}(T)}\;,
\end{equation}
where $\nabla_{\mathrm{s}}$ is the surface divergence operator, $\hat{\bm n}$ is the unit normal vector pointing from the liquid toward the vapor, $\sigma_{\mathrm{lv}}$ is the liquid-vapor surface tension at temperature $T$, and
\begin{equation}
P
:=
p_{\mathrm v}-p_{\mathrm l} 
\end{equation}
is the so-called Laplace pressure. Within each unit cell,  $p_{\mathrm v}$ and $p_{\mathrm l}$ are taken to be spatially uniform. Since the vapor-phase pressure losses are assumed negligible compared with the viscous losses in the liquid wick, $p_{\mathrm v}$ is treated as constant and equal to the reservoir pressure. The increase in interface curvature along the wick is therefore interpreted as a consequence of the decreasing liquid pressure. Note that Eq.~\eqref{eq:Young_Laplace} represents the static limit of a more general interface stress balance. A detailed discussion is provided in the Supplemental Material S2.

Dry-out occurs when the meniscus can no longer remain pinned to the pillar top and recedes toward the substrate. This depinning condition is described by a receding contact angle, $\theta_{\mathrm{r}}$ [see Fig.~\ref{fig:Array_level_sims_notation}]. $\theta_{\mathrm{r}}$ is treated as a material-system-dependent threshold. Experimental observations on heated water-filled micropillar arrays indicate that this angle is approximately independent of the applied heat flux and micropillar array geometry \cite{Antao.2016}.

For sufficiently sparse square arrays of cylindrical pillars, the corresponding critical Laplace pressure can be estimated from the analytical force-balance expression
\begin{equation}
\label{eq:Critical_pressure_sparse_array}
    P_{\mathrm{crit}} 
    = 
    \frac{4 \sigma_{\mathrm{lv}}(T) \cos\!{\left( \theta_{\mathrm{r}}\right)}}{d \left( \frac{4}{\pi} \left( \frac{l}{d}\right)^2-1\right)}\;,
\end{equation}
where $d$ is the pillar diameter, $l$ is the pitch of the square pillar lattice [see Fig.~\ref{fig:Geometry_BC_sketch}]. The liquid-vapor surface tension $\sigma_{\mathrm{lv}}$ is evaluated at the three-phase contact line (for details see Supplemental Material S2).
Dry-out is assumed to occur locally once
\begin{equation}
P
=
P_{\mathrm{crit}} \;.
\end{equation}

For liquid $\text{N}_2$, dynamic-wetting experiments on quartz yield a static contact angle of $\theta_{\mathrm s}=\qty{0}{\degree}$ and show that the receding contact angle remains close to this equilibrium value \cite{Fiorini.2023}. We therefore select $\theta_{\mathrm{r}}=\qty{0}{\degree}$ as the limiting receding angle for liquid $\text{N}_2$. Although these measurements were performed on quartz rather than silicon, the silicon surfaces considered here are expected to be covered by a native $\text{SiO}_2$ layer of order \qtyrange{1}{3}{\nano\meter} \cite{Morita.1990, Raider.1975}, making the quartz data a relevant reference.  This zero-angle limit differs notably from water-filled micropillar wicks, where receding angles are believed to be $\geq \qty{10}{\degree}$ \cite{Raj.2013, Adera.2016}. Setting $\theta_{\mathrm{r}}\rightarrow\qty{0}{\degree}$ introduces both numerical and physical challenges: the Young–Laplace solution near the contact line becomes extremely steep and ill-conditioned, while the continuum description itself becomes questionable. The zero-angle limit is therefore treated using extrapolation or clipping of the relevant quantities near complete wetting. This treatment has only a small effect on the capillary-limit estimate, because the critical pressure depends on the contact angle through $\cos(\theta_{\mathrm{r}})$ in Eq.~\eqref{eq:Critical_pressure_sparse_array}. Increasing $\theta_{\mathrm{r}}$ from $\qty{0}{\degree}$ to $\qty{10}{\degree}$ reduces $P_{\mathrm{crit}}$ by approximately $\qty{1.5}{\percent}$, since $1-\cos(\qty{10}{\degree})\approx\qty{1.5}{\percent}$.

\subsection{Thermal simulation of a unit cell}
\label{sec:thermal_simulation_unit_cell}

The steady-state temperature field is obtained from the heat-conduction equation in the absence of body heat sources or sinks
\begin{align}
\label{eq:heat_eq}
\nabla \cdot \bm{q} &= 0 \;, \\
\bm{q} &= - \kappa \nabla T \;, \label{eq:Fourier_law}
\end{align}
where $T$ denotes the temperature field, $\bm{q}$ the local heat-flux, and $\kappa$ the thermal conductivity of the respective material. In the general anisotropic case, Fourier's law [Eq.~\eqref{eq:Fourier_law}] involves the thermal-conductivity tensor. Here, liquid \ce{N2} is isotropic, therefore $\underline \kappa_{\ce{N2}} = \kappa_{\ce{N2}} \underline I$, with $\kappa_{\ce{N2}} \in \mathbb R_+$ and $\underline I$ the identity tensor. The cubic symmetry of single-crystalline \ce{Si} likewise leads to an isotropic second-rank thermal-conductivity tensor, $\underline \kappa_{\ce{Si}} = \kappa_{\ce{Si}} \underline I$ \cite{Mishra.2015}. Fourier's law can therefore be written in the scalar form given in Eq.~\eqref{eq:Fourier_law}.

The thermal boundary conditions are applied to the half unit-cell cross section shown in Fig.~\ref{fig:Geometry_BC_sketch}(c). The boundary is described by the outward unit-normal vector field $\hat{\bm n}$. At the lateral (symmetry) planes, mirror symmetry / the approximate local unit cell treatment gives a homogeneous Neumann condition,
\begin{equation}
\hat{\bm n}  \cdot \bm q = 0 \; .
\end{equation}
The exposed top surface of the silicon pillar is approximated as thermally insulating.

At the bottom surface of the \ce{Si} substrate, a spatially uniform inward heat flux $q_{\mathrm{heat}}\in\mathbb{R}_{+}$ is prescribed to represent heat dissipation within a thin active device layer,
\begin{equation}
\hat{\bm n} \cdot \bm q = -q_{\mathrm{heat}} \; .
\end{equation}

At the liquid-vapor interface, heat removal by evaporation is described using the Hertz-Knudsen-Schrage model \cite{Schrage.1953}. The net interfacial mass flux $j_{\mathrm{lv}}$ is defined positive from the liquid toward the vapor and is determined implicitly from the kinetic relation and interfacial mass conservation,
\begin{align}
\label{eq:schrage_kinetic}
j_{\mathrm{lv}} &= \sigma_{\text{ac}} \sqrt{\frac{m}{2 \pi k_{\mathrm{B}}}} \left( \frac{p_{\mathrm{sat}}(T_{\mathrm l})}{\sqrt{T_{\mathrm l}}} - \Gamma\left(\bar{v}_{\mathrm n} \beta_{\mathrm v}\right) \frac{p_{\mathrm v}}{\sqrt{T_{\mathrm v}}} \right) \\
\label{eq:schrage_mass_conservation}
j_{\mathrm{lv}} &= \rho_{\mathrm v} \bar{v}_{\mathrm n} \; ,
\end{align}
where $\sigma_{\text{ac}} \in [0,1]$ is the so-called accommodation coefficient, $m$ is the mass of a single \ce{N2} molecule, $k_{\mathrm{B}}$ is the Boltzmann constant, $T_{\mathrm l}$ is the liquid surface temperature, and $p_{\mathrm{sat}}(T_{\mathrm l})$ is its corresponding saturation vapor pressure. The far-field vapor pressure, temperature and mass density are denoted by $p_{\mathrm v}$, $T_{\mathrm v}$ and $\rho_{\mathrm v}$, respectively. The mean normal vapor drift velocity is denoted by $\bar v_{\mathrm n}$.
The predicted evaporation rate and resulting temperature field are sensitive to $\sigma_{\mathrm{ac}}$. Its estimation is discussed in Appendix~\ref{sec:accomodation_coefficient}.

The velocity-shift correction factor is defined via
\begin{equation}
\Gamma(\bar{v}_{\mathrm n} \beta_{\mathrm v}) := \exp{\left(-(\bar{v}_{\mathrm n} \beta_{\mathrm v})^2\right)} - \bar{v}_{\mathrm n} \beta_{\mathrm v} \sqrt{\pi} \left(1 - \operatorname{erf}(\bar{v}_{\mathrm n} \beta_{\mathrm v})\right) \; ,
\end{equation}
with
\begin{equation}
    \beta_{\mathrm v} := \sqrt{\frac{m}{2 k_{\mathrm{B}} T_{\mathrm v}}} \; .
\end{equation}
The full nonlinear formulation of Eq.~\eqref{eq:schrage_kinetic} is retained rather than applying the weak-evaporation linearization ($\beta_{\mathrm{v}}\bar{v}_{\mathrm n} \ll 1$). This avoids the additional approximation associated with linearizing the velocity-shift correction, which increasingly overpredicts the evaporation flux as the vapor drift velocity increases (see Supplemental Material S3 for details). The implicit system Eqs.~\eqref{eq:schrage_kinetic} and \eqref{eq:schrage_mass_conservation} is numerically solved for $j_{\mathrm{lv}}$ as a function of $T_{\mathrm l}$ for fixed far field vapor conditions. For the thermal boundary condition we convert the HKS mass flux to the corresponding heat flux 
 \begin{equation}
     q_{\mathrm{HKS}} =j_{\mathrm{lv}} h_{\mathrm{lv}} \; , 
 \end{equation}
where $h_{\mathrm{lv}}(T_{\mathrm{l}}, p_{\mathrm{l}}~\approx p_{\mathrm{v}})$ is the enthalpy of vaporization.
The thermal boundary condition for the liquid-vapor interface then has the form
\begin{equation}
\label{eq:HKS_BC}
  \hat{\bm n} \cdot \bm q = q_{\mathrm{HKS}}  \; .
\end{equation}

\subsection{Liquid flow simulation within the unit cell} \label{sec:liquid_flow_simulation_unit_cell}

The steady-state liquid-flow field in the unit cell is obtained from the steady Navier-Stokes equations with temperature-dependent material properties employing Stokes hypothesis,
\begin{align}
\label{eq:fluid_incompressibility}
    \nabla \cdot \left(\rho_{\mathrm{l}} \bm v  \right) &= 0 \; , \\
    \label{eq:Navier_Stokes}
    \rho_{\mathrm{l}} \left( \bm v \cdot \nabla \right) \bm v &= \nabla \cdot \left( - p_{\mathrm{l}} \underline I + \underline K \right) \; ,
\end{align}
where the constitutive relation for the viscous stress tensor $\underline K$ is given by
\begin{equation}
    \underline K = \mu_{\mathrm{l}} \left( \nabla \bm v + \left( \nabla \bm v \right)^{\top} - \frac{2}{3} \left( \nabla \cdot \bm v\right) \underline I \right) \; .
\end{equation}
In these expressions, $\rho_{\mathrm{l}}$ represents the liquid density, $\bm v$ is the velocity vector field, $p_{\mathrm{l}}$ is the liquid's pressure, $\mu_{\mathrm l}$ is the liquid's dynamic viscosity, and $\underline I$ denotes the identity tensor. Gravity is neglected in Eq.~\eqref{eq:Navier_Stokes}. This approximation is justified in Appendix~\ref{sec:neglect_of_gravity}.

The boundary conditions are applied to the half unit cell shown in Fig.~\ref{fig:Geometry_BC_sketch}(b).
At the silicon-liquid interface, a no-slip boundary condition is imposed, given by
\begin{equation}
\label{eq:no_slip_BC}
    \bm v = 0 \; . 
\end{equation}
At the liquid-vapor interface, an impermeable, shear-free boundary condition is used, as commonly adopted in micropillar-wick models \cite{Zhu.2016, Adera.2016, Wei.2018, Ranjan.2012, Anand.2026} (see Appendix \ref{sec:ThermalEffectsAppendix}).
This imposes zero normal velocity,
\begin{equation}
\label{eq:no_perpendicular_flow_BC}
    \hat{\bm n} \cdot \bm v = 0 \; ,
\end{equation}
and zero tangential viscous stress,
\begin{equation}
\label{eq:no_tangential_shear_stress_BC}
    \underline K \hat{\bm n} - (\underline K \hat{\bm n} \cdot \hat{\bm n}) \hat{\bm n} = 0 \; .
\end{equation}
The same boundary conditions, Eqs.~\eqref{eq:no_perpendicular_flow_BC} and \eqref{eq:no_tangential_shear_stress_BC}, are imposed at the lateral symmetry planes of the half unit cell. In this case, however, they arise due to symmetry and continuity reasons.

To evaluate the cell-level hydraulic response, uniform pressures are prescribed at the inlet and outlet planes located midway between neighboring pillars. The pressure drop is defined as
\begin{equation}
\Delta p_{\mathrm{cell}} = p_{\mathrm{l,edge},n-1} - p_{\mathrm{l,edge},n} \; ,
\end{equation}
where $p_{\mathrm{l,edge},n-1}$ and $p_{\mathrm{l,edge},n}$ denote the upstream and downstream liquid pressures, respectively [see Fig.~\ref{fig:Geometry_BC_sketch}(b)]. Because temperature-dependent density variations prevent volumetric flow conservation within the cell, the model is formulated in terms of the conserved mass flow rate $\dot{m}_{\mathrm{liq,cell}}$. The unit-cell response is characterized by a unified mass-flow resistance
\begin{equation}
\label{eq:mass_flow_resistance}
R_{\mathrm{cell}} = \frac{\Delta p_{\mathrm{cell}}}{\dot{m}_{\mathrm{liq,cell}}} \; ,
\end{equation}
which inherently accounts for the coupled variations in cross-sectional flow area, and temperature-dependent fluid properties.
Supplemental Material S5.4 shows that $\dot m_{\mathrm{liq,cell}}$ scales approximately linearly with $\Delta p_{\mathrm{cell}}$, supporting the use of a pressure-drop-independent $R_{\mathrm{cell}}$ for a fixed unit-cell state.

\section{Array-level model}

\begin{figure}[pos=h]
\centering
\includegraphics{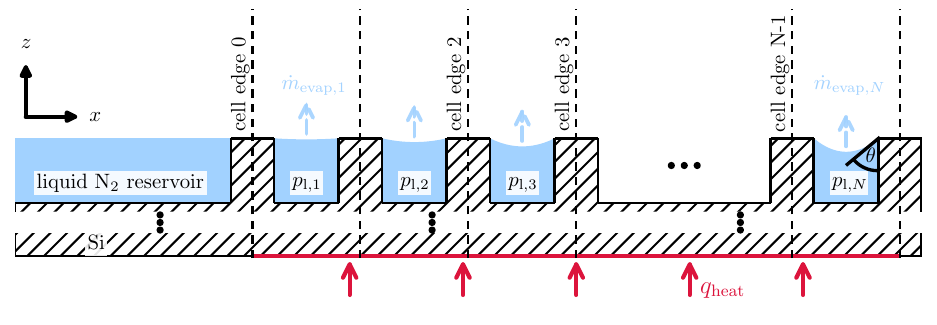}
\caption{Array-level notation for a single-fed micropillar wick along the wicking direction $x$. Cell edges are indexed from $0$ to $N$, and unit cells from $1$ to $N$. The corresponding liquid pressures are $p_{\mathrm l,\mathrm{edge},n}$ and $p_{\mathrm l,n}$. The reservoir is treated as an ideal liquid supply with negligible pressure drop.}
\label{fig:Array_level_sims_notation}
\end{figure}

In the following, we describe the discrete array model used to compute the mass-flow profile, Laplace-pressure profile, representative wall temperature, and resulting dry-out length. We consider a micropillar wick array as sketched in Figs.~\ref{fig:Geometry_BC_sketch} and \ref{fig:Array_level_sims_notation}.

\subsection{Self-consistent thermo-hydraulic solver and dry-out criterion}
\label{sec:array_pressure_mass_flow_profile}

For a candidate wick length containing $N$ unit cells in the wicking direction, the steady-state mass-flow and pressure profiles are obtained iteratively. The array model uses the Laplace pressure, $P=p_{\mathrm v}-p_{\mathrm l}$, as its local pressure variable.
Since the vapor pressure is assumed spatially uniform and equal to the reservoir pressure, an increase in $P$ along the wick directly corresponds to a decrease in liquid pressure. At the reservoir, the liquid and vapor pressures are taken to be equal, so that
\begin{equation}
P_{\mathrm{edge},0}= \qty{0}{\pascal} \;.
\end{equation}

The cell-centered Laplace pressure in cell $n\in\mathbb N_+$ is denoted by $P_n$. For a given applied heat flux $q_{\mathrm{heat}}$, the precomputed unit-cell data provide the local evaporation rate and mass-flow resistance as
\begin{equation}
\dot m_{\mathrm{evap},n}
=
\dot m_{\mathrm{evap}}
\left(
q_{\mathrm{heat}},P_n
\right),
\qquad
R_{\mathrm{cell}}
=
R_{\mathrm{cell}}
\left(
q_{\mathrm{heat}},P_n
\right) \;.
\end{equation}
The total evaporated mass flow rate over the wick is
\begin{equation}
\dot m_{\mathrm{evap,total}}
=
\sum_{n=1}^{N}
\dot m_{\mathrm{evap},n} \;.
\end{equation}
Mass conservation then gives the liquid mass flow through cell $n$ as the sum of all evaporation rates downstream of that cell,
\begin{equation}
\dot m_{\mathrm{liq},n}
=
\sum_{i=n}^{N}
\dot m_{\mathrm{evap},i} \;.
\end{equation}
Thus, the liquid mass flow is largest at the reservoir side and decreases monotonically along the wick.

The Laplace-pressure profile is updated sequentially from the reservoir using the local cell resistance $R_{\mathrm{cell}}$. The downstream edge value of cell $n$ is
\begin{equation}
P_{\mathrm{edge},n}
=
P_{\mathrm{edge},n-1}
+
R_{\mathrm{cell},n}
\dot m_{\mathrm{liq},n} \;.
\end{equation}
The updated cell-centered Laplace-pressure is approximated by the arithmetic mean of the upstream and downstream edge values,
\begin{equation}
\label{eq:def_cell_pressure}
P_n^{\mathrm{new}}
=
\frac{
P_{\mathrm{edge},n-1}
+
P_{\mathrm{edge},n}
}{2} \;.
\end{equation}

Because fluid properties are temperature-dependent, most notably the liquid-vapor surface tension $\sigma_{\mathrm{lv}}$, the array-level hydraulic model is coupled to the thermal state of the wick. In the present reduced-order treatment, the wick is assigned a representative mean wall temperature $\bar T_{\mathrm{wall}}$. The use of this mean temperature is justified a posteriori by the substrate-scale thermal analysis in Sec.~\ref{sec:array_level_temperature_averaging}, which shows that lateral heat spreading in the silicon strongly smooths the wall-temperature profile. The value of $\bar T_{\mathrm{wall}}$ is obtained from the temperature-averaging procedure described in that section [Eq.~\eqref{eq:mean_wall_temperature}] and is then used to evaluate $\sigma_{\mathrm{lv}}(\bar T_{\mathrm{wall}})$ and, consequently, the critical Laplace pressure $P_{\mathrm{crit}}$.

For each candidate wick length, the mass-flow profile, Laplace-pressure profile $P_n$, representative wall temperature $\bar T_{\mathrm{wall}}$, and critical pressure $P_{\mathrm{crit}}$ are iterated until the maximum relative change between successive iterations falls below $10^{-5}$. The dry-out condition is then evaluated.

The wick is considered stable if the local Laplace pressure remains below the critical Laplace pressure in every cell,
\begin{equation}
P_n
<
P_{\mathrm{crit}}
\qquad
\text{for all } n \;.
\end{equation}
Dry-out is assumed to occur once this condition is first violated, for at least one cell
\begin{equation}
P_n
\geq
P_{\mathrm{crit}} \; .
\end{equation}
The dry-out length is determined by increasing $N$ until this criterion is reached. The corresponding critical wick length is
\begin{equation}
L_{\mathrm{dry\text{-}out}}
=
N_{\mathrm{crit}} l \;.
\end{equation}

\section{Results and Analysis}
\label{sec:results_analysis}\subsection{Unit-cell thermal resistance and model-validity range}
\label{sec:unit_cell_thermal_res}

\begin{figure}[pos=htbp]
\centering
\includegraphics[width=1.01\linewidth]{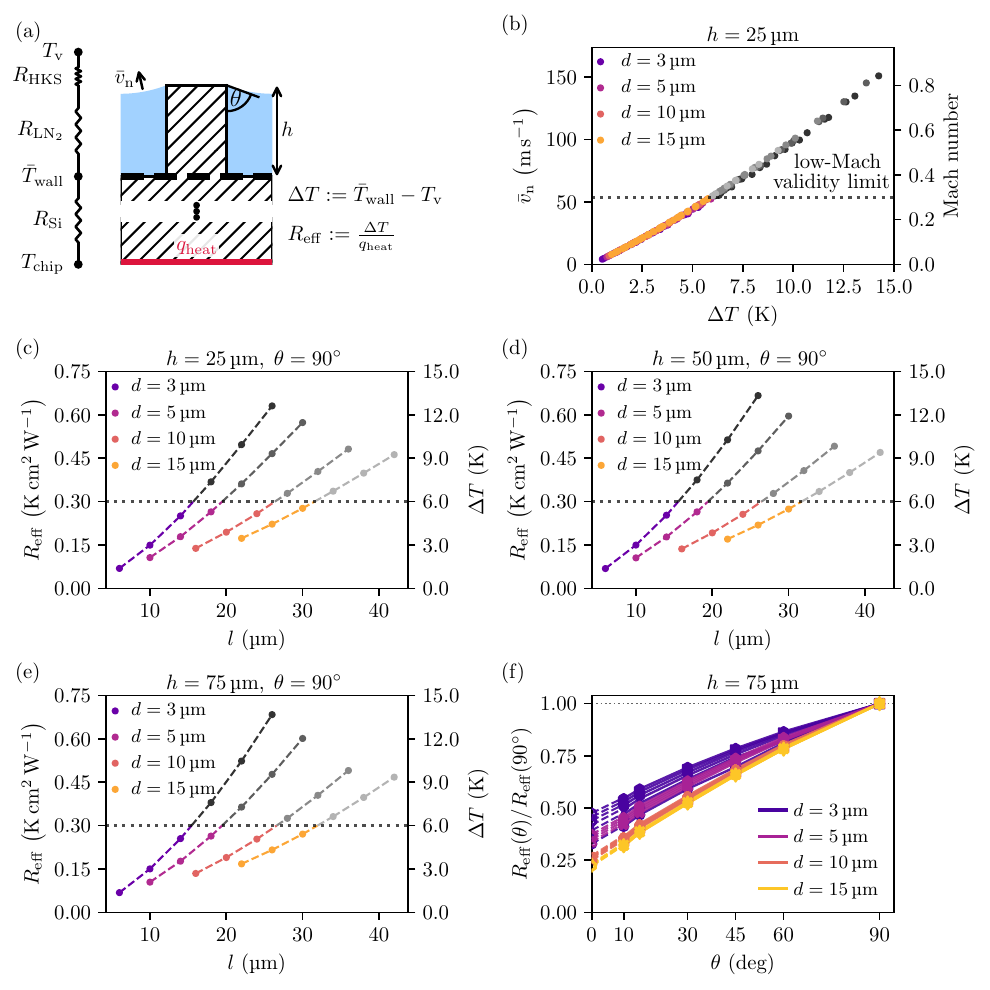}
\caption{Unit-cell thermal performance and vapor-model validity. (a) Definition and conceptual decomposition of $R_{\mathrm{eff}}$. (b) Schrage mean drift velocity versus wall superheat. The dotted line denotes the adopted low-Mach threshold. (c)-(e) $R_{\mathrm{eff}}$ versus geometry for $h=\qty{25}{\micro\meter}$, $\qty{50}{\micro\meter}$, and $\qty{75}{\micro\meter}$ at $\theta=\qty{90}{\degree}$. Grayscale data exceed the adopted validity threshold. (f) Contact-angle dependence for $h=\qty{75}{\micro\meter}$. The dashed segment denotes extrapolation from $\theta=\qty{10}{\degree}$ to $\theta=\qty{0}{\degree}$.}
\label{fig:micro_wick_ThermalRes_comparison}
\end{figure}

As illustrated by the motivating comparison in Fig.~\ref{fig:TemperatureProfileComparison}(b), the silicon domain remains nearly isothermal for the investigated geometries. This behavior follows from the large thermal-conductivity contrast at liquid-\ce{N2} temperatures: the thermal conductivity of high-purity single-crystalline silicon exceeds $\qty{1400}{\watt\per\meter\per\kelvin}$ \cite{Touloukian.1970}, whereas that of liquid \ce{N2} is approximately $\qty{0.14}{\watt\per\meter\per\kelvin}$ \cite{Ottosen.1980}. The corresponding heat-transfer path is represented schematically in Fig.~\ref{fig:micro_wick_ThermalRes_comparison}(a) by a series network comprising the silicon areal resistance $R_{\mathrm{Si}}$, an effective liquid-\ce{N2}  areal resistance $R_{\mathrm{L}\ce{N2}}$, and the kinetic liquid-vapor areal resistance $R_{\mathrm{\mathrm{HKS}}}$. An order-of-magnitude estimate gives $R_{\mathrm{Si}}\sim\qty{3.6e-3}{\kelvin\centi\meter\squared\per\watt}$ for the $\qty{500}{\micro\meter}$ silicon substrate and $R_{\mathrm{\mathrm{HKS}}}\sim\qty{1.4e-3}{\kelvin\centi\meter\squared\per\watt}$ for the liquid-vapor interface, whereas even a $\qty{1}{\micro\meter}$ characteristic conductive path through liquid \ce{N2} corresponds to $R_{\mathrm{L}\ce{N2}}\sim\qty{7.1e-2}{\kelvin\centi\meter\squared\per\watt}$. Although the liquid-side heat flow is inherently multidimensional and the series network therefore represents a scaling model rather than an exact decomposition, the comparison shows that the dominant temperature drop is expected on the liquid side. Consistently, the FEM simulations exhibit the strongest temperature gradients near the liquid-vapor interface and particularly near the three-phase contact line, where the local evaporative heat flux is highest [see Fig.~\ref{fig:3dFEM_resultimages}]. A derivation of these estimates is provided in Supplemental Material S4.

The thermal response of each unit cell is quantified by the areal effective thermal resistance
\begin{equation}
\label{eq:def_effective_areal_thermal_resistance}
R_{\mathrm{eff}}
:=
\frac{\Delta T}{q_{\mathrm{heat}}} \;,
\end{equation}
with the wall superheat
\begin{equation}
\label{eq:wall_superheat}
\Delta T
:=
\bar T_{\mathrm{wall}}-T_{\mathrm v} \;,
\end{equation}
where $\bar T_{\mathrm{wall}}$ denotes the area-averaged wall temperature and $T_{\mathrm v}$ the vapor reference temperature. The definition and schematic resistance network are illustrated in Fig.~\ref{fig:micro_wick_ThermalRes_comparison}(a).

The validity range of the evaporation model is assessed in Fig.~\ref{fig:micro_wick_ThermalRes_comparison}(b), where the normal Schrage drift velocity $\bar v_{\mathrm n}$ is plotted against the wall superheat. Despite the variation in geometry and contact angle, the investigated configurations approximately collapse onto a common trend. The wall superheat can therefore serve as a practical indicator of the vapor drift velocity and, consequently, of the model-validity range. At wall superheats above approximately $\qty{6}{\kelvin}$, the corresponding vapor Mach number reaches $M\approx0.3$. In this regime, compressibility effects and deviations from the low-Mach assumptions underlying the present vapor treatment may no longer be negligible. Data beyond this threshold are consequently shown in grayscale in Figs.~\ref{fig:micro_wick_ThermalRes_comparison}(c)--(e) and should be interpreted with reduced quantitative confidence.

They show $R_{\mathrm{eff}}$ for pillar heights of $\qty{25}{\micro\meter}$, $\qty{50}{\micro\meter}$, and $\qty{75}{\micro\meter}$ at a nominal contact angle of $\theta=\qty{90}{\degree}$. For all three heights, $R_{\mathrm{eff}}$ increases with increasing pitch-to-diameter ratio. This trend is consistent with the decreasing fraction of the unit-cell area located in the vicinity of the three-phase contact line, where the evaporative heat transport is most effective. In contrast, increasing the pillar height produces only a comparatively weak change in $R_{\mathrm{eff}}$. Within the present model, the total wetted side-wall area is therefore not the primary parameter controlling the unit-cell thermal resistance. Instead, the results indicate that thermal performance is governed mainly by the spatial extent of regions with a short conductive path through the liquid, in particular the thin-film region adjacent to the three-phase contact line.

This interpretation is further supported by the contact-angle dependence shown in Fig.~\ref{fig:micro_wick_ThermalRes_comparison}(f). Decreasing the contact angle reduces $R_{\mathrm{eff}}$, because the liquidvapor interface approaches the heated solid over a larger fraction of the unit cell and thereby shortens the characteristic conductive path through the liquid. The simulations were restricted to $\theta\geq\qty{10}{\degree}$. At smaller contact angles, the Young-Laplace solution near the contact line becomes increasingly steep and numerically ill-conditioned, while the continuum description of the immediate contact-line region eventually becomes questionable. The extrapolation toward $\theta=\qty{0}{\degree}$ should therefore be regarded only as qualitative. In particular, a vanishing $R_{\mathrm{eff}}$ is not expected, because finite thermal and kinetic resistances remain even as the liquid film becomes very thin.

The observed geometry trends are qualitatively consistent with previous studies of micropillar wicks using water as the working fluid. In particular, an increasing thermal resistance with increasing pitch to diameter ratio and a comparatively limited increase with increasing pillar height have been reported for silicon and copper micropillar structures \cite{Wei.2018, Zhang.2018, Nam.2011,Vaartstra.2019}. Likewise, reductions in thermal resistance with decreasing contact angle have been observed numerically for water-based silicon and copper micropillar systems \cite{Sharratt.2012,Vaartstra.2019}.

Having characterized $R_{\mathrm{eff}}$ for isolated unit cells, the following section extends these results to predict the overall thermal response of a full micropillar array.

\subsection{Array-level temperature averaging}
\label{sec:array_level_temperature_averaging}

\begin{figure}[pos=htbp]
\centering
\includegraphics[width=1.0\linewidth]{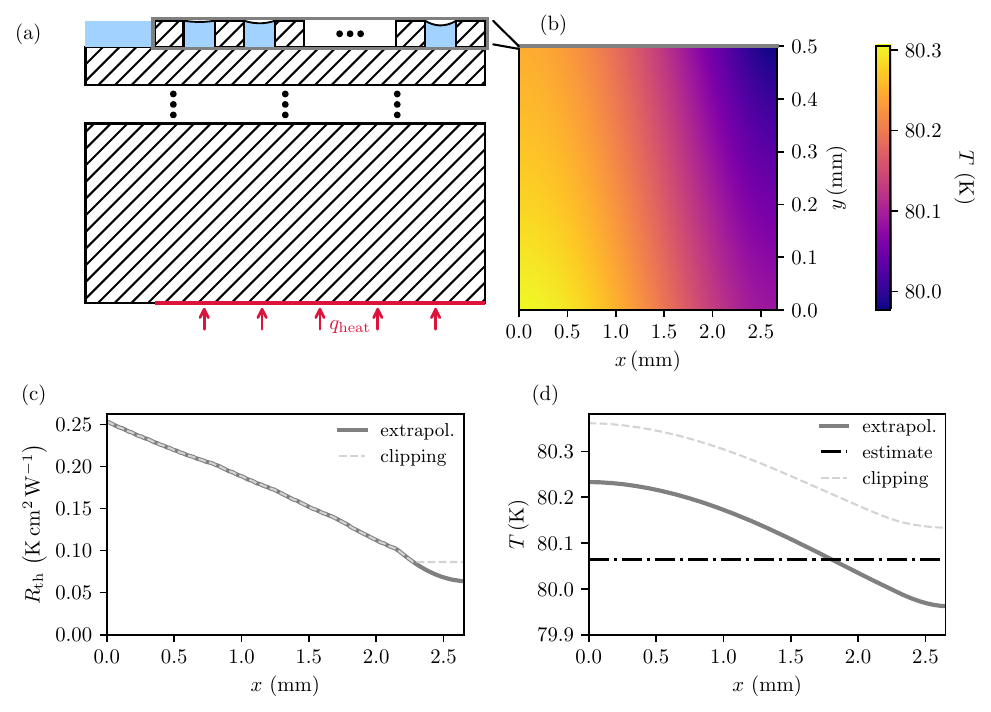}
\caption{Array-level thermal response for \(d=\qty{10}{\micro\meter}\), \(l=\qty{24}{\micro\meter}\), \(h=\qty{75}{\micro\meter}\), \(L\approx\qty{2.6}{\milli\meter}\), and \(q_{\mathrm{heat}}=\qty{20}{\watt\per\centi\meter\squared}\). (a) Schematic of the modeled micropillar array. (b) Substrate temperature field from the two-dimensional heat-conduction model. (c) Streamwise effective areal thermal resistance using extrapolation below the directly simulated range and, for comparison, clipping at \(R_{\mathrm{eff}}=\qty{0.086}{\kelvin\centi\meter\squared\per\watt}\) (corresponding to $\theta=\qty{10}{\degree}$). (d) Corresponding top-surface temperature profiles using extrapolation (solid gray) and clipping (dashed light gray), together with the analytical estimate from Eq.~\eqref{eq:mean_wall_temperature} (dash-dotted black).}
\label{fig:thermal_distribution_example}
\end{figure}

In a full micropillar wick, the local effective thermal resistance $R_{\mathrm{eff}}$ varies along the wicking direction because the streamwise liquid-pressure drop changes the capillary pressure, meniscus shape, and liquid-side heat-transfer resistance\cite{Vaartstra.2019}, see Fig.~\ref{fig:thermal_distribution_example}(a),  (c). These local variations, however, do not translate directly into comparable wall-temperature variations. The high thermal conductivity of crystalline silicon at cryogenic temperatures strongly couples neighboring unit cells. Consistent conductive smoothing has been observed in water-fed silicon micropillar evaporators, where Micro-Raman thermometry showed nearly uniform pillar temperatures and only modest streamwise temperature variations even as the meniscus evolved toward dry-out \cite{Zhang.2018}. For liquid \ce{N2}, this effect is expected to be stronger because the silicon-to-liquid thermal-conductivity contrast near $\qty{77}{\kelvin}$ exceeds that of silicon-water systems from room temperature to the boiling point by more than a factor of 40 \cite{Touloukian.1970,Ottosen.1980,Ramires.1995}.

To quantify the resulting lateral heat spreading, a two-dimensional substrate-scale heat-conduction simulation was performed for the representative array in Fig.~\ref{fig:thermal_distribution_example}. A uniform heat flux of \qty{20}{\watt\per\centi\meter\squared} was imposed at the substrate bottom, while the spatially varying $R_{\mathrm{eff}}(x)$ from the coupled wick calculation was applied at the top boundary. For local meniscus states corresponding to $\theta<\qty{10}{\degree}$, where direct unit-cell simulations are unavailable, $R_{\mathrm{eff}}$ was linearily extrapolated from the simulated range. As a sensitivity check, $R_{\mathrm{eff}}$ was instead held at its $\theta=\qty{10}{\degree}$ value of $\approx \qty{0.086}{\kelvin\centi\meter\squared\per\watt}$, see Fig.~\ref{fig:thermal_distribution_example}(c).

Despite the substantial streamwise variation in $R_{\mathrm{eff}}$, the silicon top surface remains nearly isothermal, see  Fig.~\ref{fig:thermal_distribution_example}(b),  (d). Moreover, replacing the extrapolated low-$R_{\mathrm{eff}}$ branch by the clipping treatment changes the resulting temperature profile by only about $\qty{0.1}{\kelvin}$. The predicted substrate temperature is therefore only weakly sensitive to the treatment of $R_{\mathrm{eff}}$ below the directly simulated range.

This conductive smoothing motivates a reduced-order estimate of the mean wall temperature. In the near-isothermal limit, the individual unit cells act approximately as parallel thermal conductances between a common wall temperature and the vapor. For identical unit-cell areas, the corresponding array-level areal effective resistance is therefore the harmonic mean of the local $R_{\mathrm{eff}}$ values,
\begin{equation}
\label{eq:mean_wall_temperature}
\bar T_{\mathrm{wall}}
\approx
T_{\mathrm v}
+
\frac{
q_{\mathrm{heat}}
}{
\dfrac{1}{N}
\sum_{n=1}^{N}
R_{\mathrm{eff},n}^{-1}
} \;.
\end{equation}
The derivation is given in Appendix~\ref{app:mean_wall_temperature_estimate}.

For the example in plotted Fig.~\ref{fig:thermal_distribution_example}, Eq.~\eqref{eq:mean_wall_temperature} gives $\bar T_{\mathrm{wall}}\approx\qty{80.06}{\kelvin}$. The analytical estimate (dash-dotted black) closely agrees with the substrate-scale temperature profile obtained using the extrapolated $R_{\mathrm{eff}}$ values (solid gray), as shown in Fig.~\ref{fig:thermal_distribution_example}(d). Equation~\eqref{eq:mean_wall_temperature} therefore provides a compact connection between the spatially varying unit-cell response and the array-level thermal state. In the dry-out analysis below, it is used to determine the representative wall temperature for each candidate wick length and, consequently, the surface tension entering the critical-pressure estimate.

\subsection{Capillary-limited dry-out length}
\label{sec:dry_out_results}

\begin{figure}[pos=h]
\centering
\includegraphics{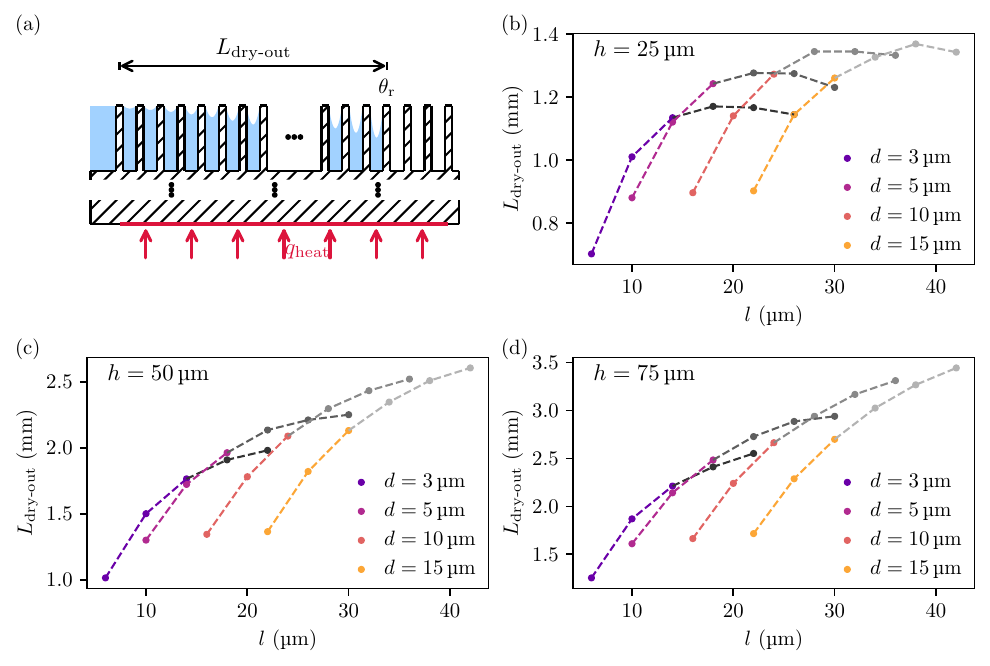}
\caption{Predicted capillary-limited dry-out length of square micropillar arrays as a function of pitch for different pillar diameters. (a) Schematic definition of the dry-out length. Predicted dry-out lengths are shown for (b) $h=\qty{25}{\micro\meter}$, (c) $h=\qty{50}{\micro\meter}$, and (d) $h=\qty{75}{\micro\meter}$. Dashed lines serve as guides to the eye. Data points for which at least one local unit cell exceeds the Schrage low-Mach validity threshold are shown in grayscale to indicate reduced predictive confidence.}
\label{fig:dry_out_results}
\end{figure}

The resulting capillary-limited dry-out lengths are shown in Fig.~\ref{fig:dry_out_results}. The predicted single-fed dry-out lengths are on the millimeter scale at $\qty{20}{\watt\per\centi\meter\squared}$.

For the $h=\qty{25}{\micro\meter}$ arrays in Fig.~\ref{fig:dry_out_results}(b), the calculations suggest a broad maximum in dry-out length at intermediate diameter-to-pitch ratios, approximately $d/l\approx0.1$--$0.4$. This trend is consistent with the competition between capillary pressure and permeability reported for water-filled silicon micropillar wicks \cite{Zhu.2016}. However, the apparent maxima lie in the reduced-confidence region identified by the Schrage low-Mach criterion and should therefore be interpreted cautiously. For $h=\qty{50}{\micro\meter}$ and $h=\qty{75}{\micro\meter}$, no distinct maximum is resolved within the investigated parameter range, although the incremental increase in dry-out length with pitch becomes progressively weaker.

The predicted liquid-\ce{N2} dry-out lengths are substantially shorter than those of comparable water-fed silicon micropillar arrays. Zhang et al.~\cite{Zhang.2018}, for example, investigated water-fed arrays with $d=\qty{20}{\micro\meter}$, $h=\qty{50}{\micro\meter}$, pitches between $\qty{30}{\micro\meter}$ and $\qty{100}{\micro\meter}$, and a wicking length of $\qty{10}{\milli\meter}$. They measured dry-out heat fluxes of approximately $\qtyrange{27}{37}{\watt\per\centi\meter\squared}$. In the present model, the representative geometry $d=\qty{10}{\micro\meter}$, $l=\qty{24}{\micro\meter}$, and $h=\qty{75}{\micro\meter}$ gives a liquid-\ce{N2} dry-out length of approximately $\qty{2.7}{\milli\meter}$ at $\qty{20}{\watt\per\centi\meter\squared}$. For the same unit-cell geometry and heat flux, but with water at $T_{\mathrm{res}}=\qty{373.15}{\kelvin}$ and $p_{\mathrm v}=\qty{101.3}{\kilo\pascal}$, the predicted dry-out length increases to approximately $\qty{18.1}{\milli\meter}$. These water-based predictions are consistent with the dry-out validation in Appendix~\ref{app:Zhu16_dry_out_validation}.

Overall, the model predictions should be interpreted with care, especially beyond the Schrage low-Mach validity threshold. The streamwise variation in thermal resistance may also produce a nonuniform evaporation profile, while neglected interfacial effects such as Marangoni stresses and evaporative normal flow could further shift the predicted dry-out limit.

The following section interprets the shorter liquid-\ce{N2} transport length using a capillary-viscous scaling argument.

\subsection{Capillary-viscous scaling and design guidelines}
\label{sec:capillary_viscous_scaling_guide}

The dry-out trends can be interpreted using an approximate capillary-viscous balance. For a single-fed wick, assuming uniform heat flux, constant permeability and thermophysical properties, Darcy-like liquid flow, and a sufficiently large number of unit cells, the dry-out length is approximated by
\begin{equation}
\label{eq:dryout_scaling_estimate}
L_{\mathrm{dry\text{-}out}}
\approx
\left(
8 \frac{q_{\mathrm{LFOM}}}{q_{\mathrm{heat}}}
\frac{\dfrac{h}{d} k_{d,l,h}}
{\dfrac{4}{\pi}\left(\dfrac{l}{d}\right)^2-1}
\cos(\theta_{\mathrm r})
\right)^{1/2} \;,
\end{equation}
with
\begin{equation}
\label{eq:LFOM_definition}
q_{\mathrm{LFOM}}
:=
\frac{\sigma_{\mathrm{lv}}\rho_{\mathrm l}h_{\mathrm{lv}}}{\mu_{\mathrm l}} \;.
\end{equation}
Here, $k_{d,l,h}$ is the intrinsic permeability of the micropillar array and has units of area. The quantity $q_{\mathrm{LFOM}}$ has units of heat flux and corresponds to the liquid figure of merit for capillary-driven heat transport \cite{NASA.1972}. Equation~\eqref{eq:dryout_scaling_estimate} neglects spatial variations in temperature, thermophysical properties, permeability, and meniscus shape and should therefore be interpreted as a scaling relation rather than a quantitative replacement for the coupled array model. A derivation is provided in Appendix~\ref{app:dryout_scaling_derivation}.

Equation~\eqref{eq:dryout_scaling_estimate} separates the principal contributions to the capillary transport limit. Larger surface tension, liquid density, latent heat, permeability, and pillar height increase $L_{\mathrm{dry\text{-}out}}$, whereas larger viscosity and applied heat flux decrease it. The receding contact angle enters through $\cos(\theta_{\mathrm r})$. Geometry affects the dry-out length through both the permeability $k_{d,l,h}$ and the capillary-pressure factor determined by $h/d$ and $l/d$.

For fixed geometry and heat flux, Eq.~\eqref{eq:dryout_scaling_estimate} reduces to a comparison of fluid properties and wetting. Using representative saturation properties for liquid \ce{N2} at \qty{77}{\kelvin} and water at \qty{373}{\kelvin}, the predicted ratio of dry-out lengths is
\begin{equation}
\label{eq:dry_out_length_ratio_simple_equation}
\frac{L_{\mathrm{dry\text{-}out}}^{\ce{N2}}}
{L_{\mathrm{dry\text{-}out}}^{\ce{H2O}}}
\approx
\left(
\frac{q_{\mathrm{LFOM}}^{\ce{N2}}}
{q_{\mathrm{LFOM}}^{\ce{H2O}}}
\frac{
\cos\!\left(\theta_{\mathrm r}^{\ce{N2}}\right)
}{
\cos\!\left(\theta_{\mathrm r}^{\ce{H2O}}\right)
}
\right)^{1/2}
\approx
0.16 \;.
\end{equation}
For the representative geometry, the detailed simulations give
\begin{equation}
\label{eq:dry_out_length_ratio_FEM}
\frac{
L_{\mathrm{dry\text{-}out}}^{\ce{N2}}
}{
L_{\mathrm{dry\text{-}out}}^{\ce{H2O}}
}
\approx
\frac{\qty{2.7}{\milli\meter}}
{\qty{18.1}{\milli\meter}}
\approx
0.15 \;.
\end{equation}
The close agreement indicates that the capillary-viscous scaling captures the dominant difference between the predicted liquid-\ce{N2} and water transport lengths for this geometry.

As a second, order-of-magnitude comparison, we relate the liquid-\ce{N2} simulations for cylindrical micropillar arrays with $d=\qty{15}{\micro\meter}$, $l=\qty{34}{\micro\meter}$, and $h=\qty{25}{\micro\meter}$ at \qty{77.36}{\kelvin} and \qty{101.3}{\kilo\pascal} to the experiments of Hasan et al.~\cite{Hasan.2025}. Their structures consisted of square-base pillars with characteristic dimensions $d=\qty{400}{\micro\meter}$, $l=\qty{900}{\micro\meter}$, and $h=\qty{600}{\micro\meter}$ and were operated at approximately \qty{110}{\kelvin} and \qty{1.38}{\mega\pascal}. The pillar cross sections, aspect ratios, feeding configurations, and operating conditions are therefore not identical, and the comparison is not intended as a direct validation of the present model.
To estimate the expected scale of the critical heat flux, we nevertheless apply Eq.~\eqref{eq:dryout_scaling_estimate} using the corresponding geometric and thermophysical scaling. Under geometrically similar scaling, the intrinsic permeability scales with the square of the characteristic length, $k\propto\lambda^2$. Taking a characteristic dimensional scaling factor of approximately $\lambda=26.67$ and accounting for the different liquid-\ce{N2} properties at the two operating conditions \cite{Lemmon.1994, Lemmon.2004, Span.2000} yields an estimated critical heat flux of approximately \qty{45}{\watt\per\centi\meter\squared} when the present single-fed calculation is mapped to the experimentally supplied wick dimensions and feeding length. This value lies within the experimental range of approximately \qtyrange{37}{50}{\watt\per\centi\meter\squared} reported by Hasan et al.~\cite{Hasan.2025}. Given the differences in pillar shape, aspect ratio, and reservoir configuration, this agreement should be regarded as an order-of-magnitude consistency check rather than a quantitative validation. Determination of the absolute permeability for a specific geometry still requires numerical simulation or an appropriate analytical model.

Across working fluids, capillary transport is governed by $q_{\mathrm{LFOM}}=\sigma_{\mathrm{lv}}\rho_{\mathrm l}h_{\mathrm{lv}}/\mu_{\mathrm l}$ together with wetting and wick geometry. For liquid \ce{N2}, the low surface tension and latent heat outweigh the benefit of its low viscosity, resulting in a lower $q_{\mathrm{LFOM}}$ and shorter dry-out lengths than for water. More generally, this behavior reflects the weaker intermolecular cohesion of many cryogenic fluids, which contributes to their comparatively low surface tension and latent heat of vaporization. Equation~\eqref{eq:dryout_scaling_estimate} therefore provides a compact framework for relating these fluid-property differences to capillary-transport limits across working fluids and wick geometries.

From a geometry perspective, extending the dry-out length requires reducing hydraulic resistance while retaining sufficient capillary pressure and effective thin-film evaporation area. Potential strategies include multi-sided liquid feeding, graded or variable-permeability wick architectures \cite{Wilke.2018,Liu.2022}, and non-cylindrical pillar geometries that modify permeability, capillary pressure, and contact-line length simultaneously \cite{Ranjan.2012,Yuncu.2023,Anand.2026}. These approaches relax the single-fed, uniform square-array assumptions underlying Eq.~\eqref{eq:dryout_scaling_estimate} and indicate that the millimeter-scale lengths predicted in Fig.~\ref{fig:dry_out_results} represent a baseline for the geometries investigated here rather than a general upper limit for cryogenic micropillar wicks.

\section{Conclusion}

We numerically investigated liquid-\ce{N2}-filled silicon micropillar wicks as a general capillary-fed thin-film evaporation concept for cryogenic electronics. For the representative array at $\qty{20}{\watt\per\centi\meter\squared}$, the predicted chip-temperature rise is approximately $\qty{2.7}{\kelvin}$. This is substantially below the estimates for the representative conduction-cooled and direct-immersion benchmarks. The predicted temperature rise depends sensitively on the accommodation coefficient in the Schrage model. We estimate this coefficient using a transition-state model, as experimentally determined values for liquid \ce{N2} remain unavailable. Dedicated measurements would therefore be valuable for validating the predicted thermal performance.

High thermal conduction in the silicon substrate strongly smoothes cell-scale variations in effective thermal resistance. This allows the Si within the array to be modeled using a single spatially averaged temperature.

System scalability is primarily limited by capillary-fed dry-out. Using liquid \ce{N2} at a heat flux of $\qty{20}{\watt\per\centi\meter\squared}$, the maximum predicted single-fed length is $\qty{2.7}{\milli\meter}$, which corresponds to a maximum coolable width of $\qty{5.3}{\milli\meter}$ in an ideal double-fed layout. For the same representative geometry and heat flux, substituting water at $\qty{373}{\kelvin}$ increases this dry-out length to approximately $\qty{18.1}{\milli\meter}$. This strong dependence on the working fluid is captured by the liquid figure of merit, $q_{\mathrm{LFOM}}=\sigma_{\mathrm{lv}}\rho_{\mathrm l}h_{\mathrm{lv}}/\mu_{\mathrm l}$. When including wetting effects, this scaling approach predicts a liquid-\ce{N2}/water ratio of 0.16, showing excellent agreement with the 0.15 ratio derived from detailed simulations.

The resulting design requirement is therefore multiscale: the unit cell must provide low thermal resistance, while the array must provide sufficient capillary pressure and permeability to sustain liquid replenishment over the required lateral distance. Multi-sided feeding, graded permeability, and non-cylindrical pillar geometries therefore provide direct routes for extending the capillary-limited dimensions beyond those of the single-fed uniform square arrays considered here.

The present model assumes a quasi-static liquid--vapor interface and neglects Marangoni stresses, the normal liquid velocity associated with evaporation, and possible streamwise variations in the evaporation rate arising from the spatially varying thermal resistance. These approximations limit quantitative accuracy at high local heat fluxes and large interfacial temperature gradients. Further work should therefore combine fully coupled interfacial thermal-hydraulic simulations with experiments on micrometer-scale liquid-\ce{N2} wicks near $\qty{77}{\kelvin}$.

\section*{Declaration of competing interest}

The authors declare that they have no known competing financial interests
or personal relationships that could have appeared to influence the work
reported in this paper.

\section*{Acknowledgments}
This work was supported by JST ALCA-Next (Grant No.~JPMJAN24E2) and JST ASPIRE (Grant No.~JPMJAP2411).

\section*{Data availability}

The data that support the findings of this study are available from the
corresponding author upon reasonable request.

\appendix

\section{Representative 3D FEM fields}
\label{sec:3d_fem_fields}

\begin{figure}[pos=h]
\centering
\includegraphics{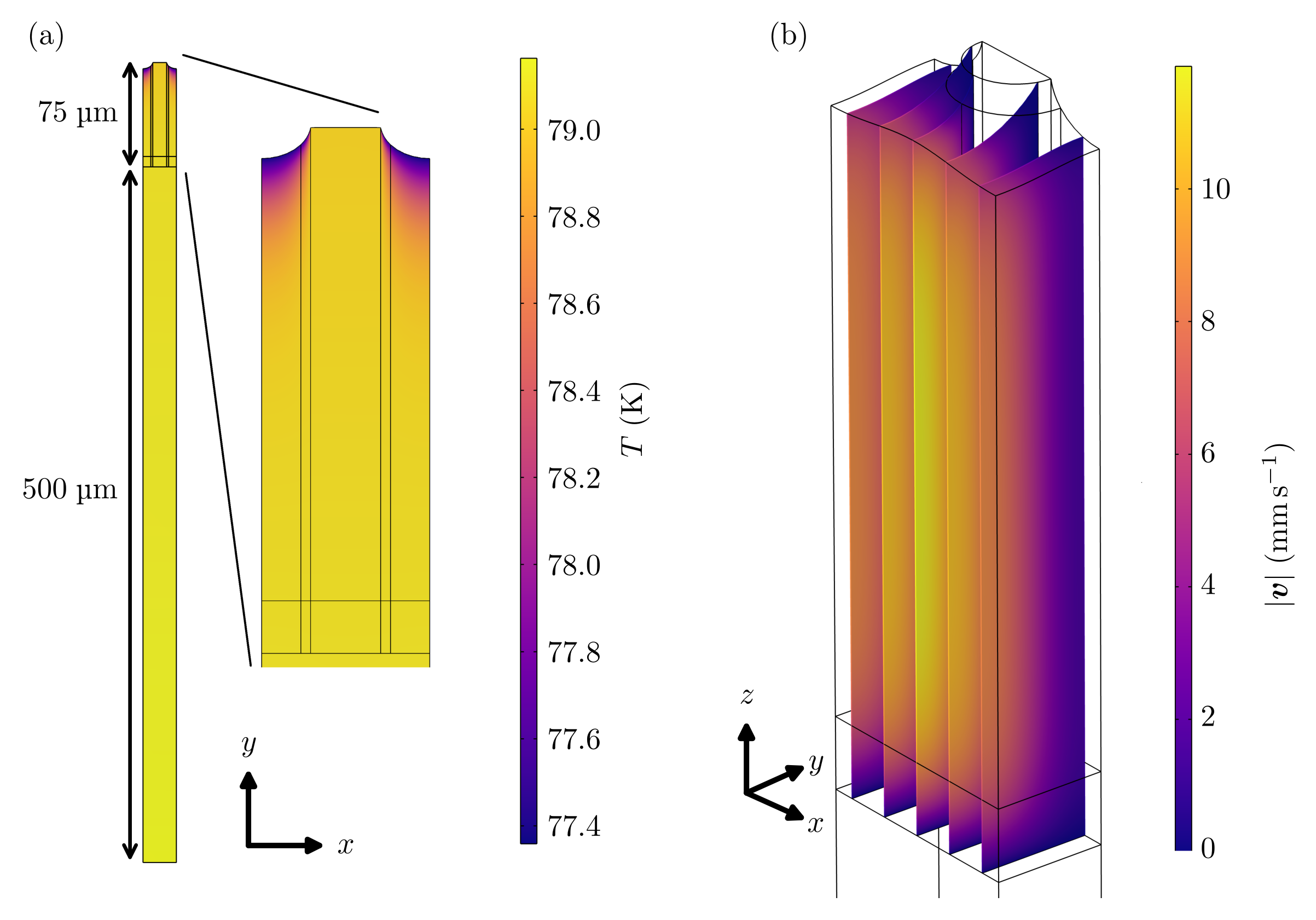}
\caption{Representative FEM fields for the unit cell with $d=\qty{10}{\micro\meter}$, $l=\qty{24}{\micro\meter}$, $h=\qty{75}{\micro\meter}$ and $\theta=\qty{10}{\degree}$. (a) Temperature field at an applied uniform heat flux of $\qty{20}{\watt\per\centi\meter\squared}$. The silicon region is nearly isothermal, while strong temperature gradients occur near the liquid-vapor interface. (b) Liquid-flow field computed for the same temperature field with an imposed pressure drop of $\Delta p_{\mathrm{cell}}=\qty{1}{\pascal}$. The flow remains smooth, with increased velocity magnitudes near the narrow flow constrictions.}
\label{fig:3dFEM_resultimages}
\end{figure}

Figure~\ref{fig:3dFEM_resultimages} shows representative three-dimensional FEM solutions for the unit-cell geometry with $d=\qty{10}{\micro\meter}$, $l=\qty{24}{\micro\meter}$, $h=\qty{75}{\micro\meter}$ and $\theta=\qty{10}{\degree}$. The region near the liquid-vapor interface and the pillar wall was resolved using a locally refined, exponentially graded mesh. Black outlines are overlaid to indicate selected simulation domain boundaries.

\section{Liquid-\ce{N2} critical heat flux calculation}
\label{sec:CHF_details}
For an extended, flat, horizontal surface cooled by saturated liquid \ce{N2} under normal boiling conditions (\qty{77.36}{\kelvin}, \qty{101.3}{\kilo\pascal}), the cryogenic CHF correlation reported by Patel et al.~\cite{Patel.2022} 
is, to a very good approximation, equivalent to the classical Kutateladze relation \cite{Kutateladze.1948, Kutateladze.1961}:
\begin{equation}
\label{eq:Kutateladze_eq}
q_{\mathrm{CHF}}
=
C_{\mathrm{K}} \rho_{\mathrm{v}} h_{\mathrm{lv}}
\left(\frac{\sigma_{\mathrm{lv}} g \left(\rho_{\mathrm l}-\rho_{\mathrm{v}}\right)
}{
\rho_{\mathrm{v}}^{2}
}
\right)^{\frac{1}{4}} \;,
\end{equation}
where $C_{\mathrm{K}}=0.16$ is the Kutateladze coefficient, $\rho_{\mathrm l}$ and $\rho_{\mathrm{v}}$ are the saturated-liquid and saturated-vapor mass densities, respectively, $h_{\mathrm{lv}}$ is the specific enthalpy of vaporization, $\sigma_{\mathrm{lv}}$ is the liquid-vapor surface tension, and $g$ is the gravitational acceleration. Liquid-\ce{N2} thermophysical properties are taken from Ottosen and Mannov~\cite{Ottosen.1980}, while the saturated-vapor density is estimated using the ideal gas law.

\section{Details of the accommodation coefficient used in the Schrage equation} \label{sec:accomodation_coefficient}

The Hertz-Knudsen-Schrage formulation in Eqs.~\eqref{eq:schrage_kinetic} and \eqref{eq:schrage_mass_conservation} contains separate coefficients for evaporation and condensation to account for non-ideal molecular transfer across the liquid-vapor interface \cite{Vaartstra.2022}. These quantities are denoted by the evaporation coefficient $\sigma_{\mathrm e}$ and the condensation coefficient $\sigma_{\mathrm c}$, respectively. Their experimental determination is difficult, and the reported values depend strongly on the measurement method and interfacial conditions.
A common simplification is to assume equal evaporation and condensation coefficients \cite{Vaartstra.2022},
\begin{equation}
\sigma_{\mathrm e} = \sigma_{\mathrm c}
=: \sigma_{\mathrm{ac}}, 
\qquad \sigma_{\mathrm{ac}}\in[0,1]\;,
\end{equation}
where $\sigma_{\mathrm{ac}}$ denotes the mass accommodation coefficient. The assumption $\sigma_{\mathrm e}=\sigma_{\mathrm c}$ is consistent with detailed balance at thermodynamic equilibrium. Its validity under strongly non-equilibrium conditions, however, is not guaranteed.

Reliable experimentally determined accommodation coefficients for \ce{N2} could not be identified. Bellur et al.~\cite{Bellur.2023} determined mass accommodation coefficients for cryogenic hydrogen and methane by combining phase-change experiments with a multiscale thermal model that accounts for non-uniform interfacial temperatures. Their results were consistent with the generalized transition-state theory proposed by Nagayama and Tsuruta \cite{Nagayama.2003}. Bellur et al.~also found broadly similar agreement when comparing the model with previously reported data for several other fluids. Motivated by this experimental evidence, the transition-state expression is adopted here to estimate the accommodation coefficient of \ce{N2}
\begin{equation}
\sigma_{\mathrm{ac}}
=\left(1-l\right)
\exp\left(-\frac{l}{2\left(1-l\right)}\right) \;,
\end{equation}
where
\begin{equation}
l:=\left(\frac{\rho_{\mathrm v}}{\rho_{\mathrm l}}\right)^{1/3} \;.
\end{equation}
Here, $\rho_{\mathrm l}$ and $\rho_{\mathrm v}$ denote the liquid- and vapor-phase mass densities, respectively. Temperature dependent liquid \ce{N2} mass densities have been sourced from Ref.~\cite{Ottosen.1980} and vapor mass densities are derived from the ideal gas law.
For \ce{N2}, the resulting accommodation coefficient exhibits only a weak temperature dependence, decreasing from approximately $\sigma_{\mathrm{ac.}} = 0.744$ at \qty{68}{\kelvin} to $\sigma_{\mathrm{ac.}} = 0.734$ at \qty{88}{\kelvin}.

In contrast to this physically motivated estimate, several macroscopic cryogenic-tank simulations employ very small accommodation coefficients, typically $\sigma_{\mathrm{ac}}=10^{-4}$--$10^{-2}$, for liquid-\ce{N2} and liquid-\ce{H2} \cite{Kassemi.2016, Ciccotosto.2021, Stewart.2015, Agui.2015}. These values are far below both the transition-state-theory estimates for the corresponding liquid-vapor density ratios and recently measured values for liquid-\ce{H2} \cite{Bellur.2023}. Some of these references additionally assume equal liquid and vapor temperatures at the interface, thereby nonphysically suppressing an important contribution to the kinetic driving force for phase change \cite{Kassemi.2016, Stewart.2015, Ciccotosto.2021}. Moreover, Refs.~\cite{Kassemi.2016, Agui.2015} found their numerics to be nearly insensitive to $\sigma_{\mathrm{ac}}$ over  multiple orders of magnitude and employed small values primarily for numerical stability reasons.  These values should therefore be regarded primarily as numerical fitting parameters rather than physically determined accommodation coefficients. If the Schrage formulation with $\sigma_{\mathrm e}=\sigma_{\mathrm c}$ is accepted, transition-state theory predicts $\sigma_{\mathrm{ac}}>0.4$ for liquid-\ce{N2} even at \qty{1}{\mega\pascal}, which exceeds the pressure range of the cited studies. These substantially smaller values for nitrogen's accommodation coefficient are therefore not adopted in the present work.

\section{Comment on approximations in the fluid flow modeling}

\subsection{Neglecting gravitational effects}
\label{sec:neglect_of_gravity}

Gravity is omitted from the fluid flow governing equations [Eq.~\eqref{eq:fluid_incompressibility}] because capillary forces dominate gravity across all investigated scales.

At the micro-structure scale, the Bond number quantifies the relative magnitude of gravitational to capillary forces
\begin{equation}
\mathrm{Bo} = \frac{(\rho_{\mathrm{l}} - \rho_{\mathrm{v}}) g h^2}{\sigma_{\mathrm{lv}}} \le 0.005 \;,
\end{equation}
where $\rho_{\mathrm{l}}$ and $\rho_{\mathrm{v}}$ denote the liquid and vapor mass densities of nitrogen, $g$ is the gravitational acceleration, $h$ is the pillar height, and $\sigma_{\mathrm{lv}}$ is the liquid-vapor surface tension.

At the macro-wick scale, the gravitational impact depends on the orientation relative to Earth's gravitational field. Comparing the hydrostatic pressure to the critical capillary pressure yields the dimensionless ratio:
\begin{equation}
\eta_{\mathrm{g}} = \frac{\rho_{\mathrm{l}} g L \cos\alpha}{P_{\mathrm{crit}}}\;,
\end{equation}
where $L$ is the total wick length and $\alpha$ is the angle between the wicking direction and the gravitational vector. Within the valid low-Mach regime, $\eta_{\mathrm{g}}$ ranges between $0.019$ and $0.084$ under the worst-case scenario of parallel alignment ($\alpha = 0^\circ$).

Thus, gravity introduces at most a minor correction to the array-scale model in parallel alignment, slightly enhancing or reducing performance depending on the flow direction, and is entirely negligible for perpendicular orientations ($\alpha = 90^\circ$). Overall, gravitational effects remain negligible for the present single-fed wicks.

\subsection{A short note on thermally induced effects at the liquid-vapor interface}
\label{sec:ThermalEffectsAppendix}
In the fluid-flow simulations, the liquid-vapor interface is modeled using an impermeable, shear-free boundary condition, as commonly adopted in micropillar-wick models \cite{Zhu.2016, Adera.2016, Wei.2018, Ranjan.2012, Anand.2026}. This approximation imposes zero normal liquid velocity and zero tangential viscous stress at the interface. Models employing this simplification have reproduced experimentally measured dry-out heat fluxes and wick lengths with reasonable accuracy \cite{Zhu.2016,Adera.2016,Wei.2018}.

Strictly, however, the no-penetration condition in Eq.~\eqref{eq:no_perpendicular_flow_BC} is incompatible with nonzero evaporation. For a stationary interface, interfacial mass conservation requires
\begin{equation}
\hat{\bm n} \cdot \bm v_{\mathrm l}
=
\frac{j_{\mathrm{lv}}}{\rho_{\mathrm l}}\;,
\end{equation}
where $\hat{\bm n}$ points from the liquid toward the vapor, and $j_{\mathrm{lv}}>0$ denotes the evaporative mass flux. The no-penetration approximation therefore neglects the normal liquid velocity associated with evaporative mass transfer.

Temperature gradients along the interface may additionally generate thermocapillary, or Marangoni, stresses. Because the surface tension of \ce{N2} decreases with increasing temperature, these stresses drive interfacial liquid flow from hotter regions of lower surface tension toward colder regions of higher surface tension. Yuncu et al.~\cite{Yuncu.2023} incorporated this coupling for water-filled silicon micropillar wicks and demonstrated significant Marangoni-induced modifications of the flow and thermal fields, although the meniscus shape was determined under quasi-static flow conditions. Their model was compared with the experimental dry-out data reported in Refs.~\cite{Zhu.2016,Adera.2016,Wei.2018}. They considered an isothermal model, a non-isothermal model incorporating temperature-dependent thermophysical properties (the present publications approach), and a non-isothermal model additionally including Marangoni stresses. For quantitatively capturing the effect of thermocapillary effects they defined the Marangoni number
\begin{equation}
\label{eq:def_Marangoni_number}
\mathrm{Ma}
:=
\frac{
\left|
\dfrac{\partial\sigma_{\mathrm{lv}}}{\partial T}
\right|
L_{\mathrm{char}}\Delta T
}{
\mu_{\mathrm l}\alpha_{\mathrm l}
}\;,
\end{equation}
where $\sigma_{\mathrm{lv}}$ is the liquid-vapor surface tension, $L_{\mathrm{char}}$ is the distance along the interface from the pillar edge to the lowest point of the meniscus between neighboring pillars, and
\begin{equation}
\Delta T
:\approx
\left|
\bar T_{\mathrm{wall}}-T_{\mathrm{v}}
\right|
\end{equation}
is the corresponding interfacial temperature difference. The quantities $\mu_{\mathrm l}$ and $\alpha_{\mathrm l}$ denote the liquid dynamic viscosity and thermal diffusivity, respectively.

Yuncu et al. reported Marangoni numbers up to approximately $\mathrm{Ma}=293$. Including temperature-dependent material properties improved the predicted dry-out heat fluxes relative to the isothermal model. Adding Marangoni stresses produced only small changes in the predicted dry-out heat flux, below about $\qty{2}{\percent}$ in their cases. The effect on wall superheat was more sensitive. For $\mathrm{Ma}<100$, the non-isothermal models with and without Marangoni stresses differed by less than about $\qty{2}{\percent}$. At the largest Marangoni numbers, however, including Marangoni stresses reduced the predicted wall superheat by about $\qty{14}{\percent}$ and improved agreement with the experimental temperature data.

We evaluate Eq.~\eqref{eq:def_Marangoni_number} for liquid \ce{N2} using material properties from Ref.~\cite{Ottosen.1980}. For the best-performing micropillar wicks that remain within the Schrage low-Mach criterion, the estimated range is $\mathrm{Ma}\approx322$--$409$. These values are comparable to, but somewhat larger than, the largest values considered by Yuncu et al. Direct quantitative transfer from the water-wick study is therefore not justified. Nevertheless, their results suggest that Marangoni stresses may have a stronger influence on the predicted wall superheat than on the capillary dry-out heat flux.

The present model therefore likely captures the leading capillary-viscous dry-out trends, but may overestimate the wall temperature when thermocapillary stresses are significant. This limitation becomes more important outside the low-Mach validity range, where larger temperature gradients and stronger interfacial nonequilibrium are expected. A fully coupled treatment of interface shape, evaporation, liquid flow, and thermocapillary stresses is required to quantify this effect for liquid-\ce{N2} micropillar wicks.

\section{Reduced-order estimate of the mean wall temperature}
\label{app:mean_wall_temperature_estimate}

We want to derive an estimate for the mean wall temperature used in Eq.~\eqref{eq:mean_wall_temperature}. The derivation assumes that lateral heat spreading in the silicon substrate is sufficiently strong that the wall temperature can be approximated by a spatially uniform value, $\bar T_{\mathrm{wall}}$. This approximation is motivated by the high thermal conductivity of silicon at liquid-\ce{N2} temperatures and by the substrate-scale temperature field shown in Fig.~\ref{fig:thermal_distribution_example}.

For a wick consisting of $N$ unit cells with identical projected footprint area $l^2$, the local effective thermal resistance of cell $n$ is denoted by $R_{\mathrm{eff},n}$. If the vapor reference temperature is spatially uniform and equal to $T_{\mathrm v}$, the integrated through-plane heat flow through cell $n$ can be approximated as
\begin{equation}
\label{eq:app_unit_cell_heat_flow}
Q_n
\approx
l^2
\frac{
\bar T_{\mathrm{wall}} - T_{\mathrm v}
}{
R_{\mathrm{eff},n}
} \; .
\end{equation}
Here, $Q_n$ is the integrated heat flow through cell $n$, whereas $R_{\mathrm{eff},n}$ is the areal thermal resistance obtained from the corresponding unit-cell simulation.

For a spatially uniform applied heat flux $q_{\mathrm{heat}}$, global energy conservation requires
\begin{equation}
\label{eq:app_global_energy_balance}
Nl^2 q_{\mathrm{heat}}
=
\sum_{n=1}^{N} Q_n \;.
\end{equation}
Substitution of Eq.~\eqref{eq:app_unit_cell_heat_flow} into Eq.~\eqref{eq:app_global_energy_balance} gives
\begin{equation}
\label{eq:app_energy_balance_substituted}
Nl^2 q_{\mathrm{heat}}
\approx
l^2
\left(
\bar T_{\mathrm{wall}} - T_{\mathrm v}
\right)
\sum_{n=1}^{N}
R_{\mathrm{eff},n}^{-1} \;.
\end{equation}
Solving for $\bar T_{\mathrm{wall}}$ yields
\begin{equation}
\label{eq:app_mean_wall_temperature}
\bar T_{\mathrm{wall}}
=
T_{\mathrm v}
+
\frac{
q_{\mathrm{heat}}
}{
\dfrac{1}{N}
\sum_{n=1}^{N}
R_{\mathrm{eff},n}^{-1}
} \;.
\end{equation}
Thus, under the assumptions of equal unit-cell footprint area, uniform applied heat flux, and nearly uniform wall temperature, the array-level effective thermal resistance is the harmonic mean of the local areal unit-cell resistances.

For arrays with spatially varying pitch, non-uniform heat input, or weaker lateral temperature homogenization in the substrate, Eq.~\eqref{eq:app_mean_wall_temperature} should be replaced by an area-weighted or fully coupled substrate-scale heat-conduction calculation.

\section{Validation of the thermal model against Adera et al.}
\label{app:thermal_model_validation_adera}

\begin{figure}[pos=htbp]
\centering
\includegraphics{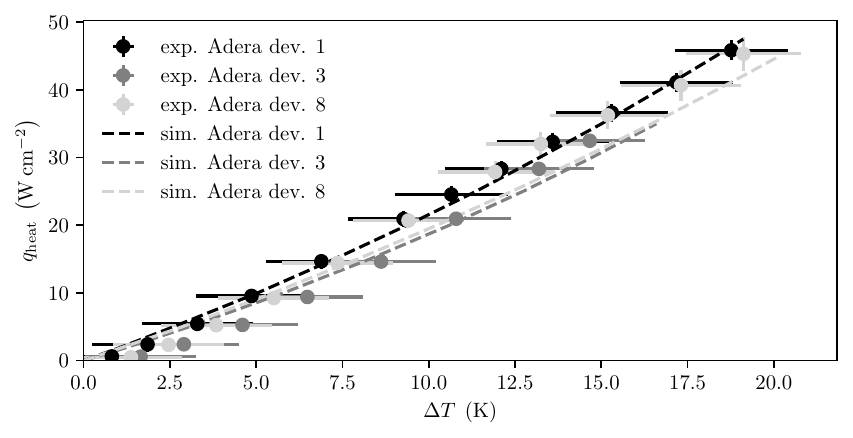}
\caption{Comparison between measured dry-out heat fluxes for devices 1, 3, and 8 of Adera et al.~\cite{Adera.2016} and the corresponding simulations obtained with the thermal-hydraulic model used in the present work. The experiments used square micropillar arrays supplied by four symmetrically arranged liquid reservoirs, whereas the simulations represent a mirror-symmetric, two-sided reservoir-fed configuration. The simulations use an accommodation coefficient of $\sigma_{\mathrm{ac}}=0.054$, a reservoir-liquid temperature of $\qty{297.23}{\kelvin}$, and a vapor pressure of $\qty{3}{\kilo\pascal}$.}
\label{fig:Adera16_comparison}
\end{figure}

As a consistency assessment, the present model was applied to the water-filled micropillar-wick experiments of Adera et al.~\cite{Adera.2016}, see Fig.~\ref{fig:Adera16_comparison}. Their interferometric measurements showed liquid-vapor interface profiles consistent with the Young-Laplace equation and yielded mean apparent contact angles of $\qty{65 \pm 4}{\degree}$ along the diagonal direction and $\qty{75 \pm 3}{\degree}$ along the lateral direction. These measurements provide an experimental reference for the contact-angle range used in the simulations.

A single accommodation coefficient, $\sigma_{\mathrm{ac}}=0.054$, was used for all eight simulated devices. For each device, the receding contact angle was varied to reproduce the measured dry-out heat flux. The resulting angles are summarized in Table~\ref{tab:Adera_contact_angles}. They range from $\qty{57.5}{\degree}$ to $\qty{72.0}{\degree}$, with a mean and sample standard deviation of $\qty{66.4 \pm 5.2}{\degree}$, well within the two standard deviations of the reported experimental uncertainties.

\begin{table}[pos=htbp]
\centering
\caption{Receding contact angles required to reproduce the measured dry-out heat fluxes of devices 1--8 reported by Adera et al.~\cite{Adera.2016}, using a common accommodation coefficient of $\sigma_{\mathrm{ac}}=0.054$.}
\label{tab:Adera_contact_angles}
\begin{tabular}{c c c c c c c c c}
\toprule
Device & 1 & 2 & 3 & 4 & 5 & 6 & 7 & 8 \\
\midrule
simulated $\theta_{\mathrm r}\;(\si{\degree})$   &
60.1 & 71.6 & 67.6 & 67.5 & 66.0 & 57.5 & 68.8 & 72.0 \\
\bottomrule
\end{tabular}
\end{table}

The accommodation coefficient used here is close to the value $\sigma_{\mathrm{ac}}=0.052$ employed by Vaartstra et al.~\cite{Vaartstra.2019} for simulations of a subset of the same experimental data.

The comparison is not fully one-to-one because the numerical model represents a two-sided, mirror-symmetric feeding configuration, whereas the experiments used four-sided reservoir feeding around a square micropillar array. Some deviation between experiment and simulation is therefore expected. Nevertheless, for the relatively large contact angles observed by Adera et al., the local thermal-response variation across the wick remains moderate, and the streamwise profiles from the reservoir edge toward the array center are expected to be qualitatively similar. The comparison is therefore useful as a consistency check for the thermal model, rather than as a strict geometry-resolved validation of the full experimental configuration.

The predicted relation between applied heat flux and wall superheat is sensitive to the accommodation coefficient used in the Hertz-Knudsen-Schrage boundary condition. Over the investigated range, this relation is approximately linear, and its slope can be adjusted substantially through $\sigma_{\mathrm{ac}}$. This sensitivity implies that micropillar-wick experiments can, in principle, provide a useful route for estimating effective accommodation coefficients, provided that the interface shape, contact angle, and thermal boundary conditions are independently constrained. At the same time, $\sigma_{\mathrm{ac}}$ can also compensate for deficiencies in the thermal or interfacial model if these quantities are not resolved accurately. The present comparison is therefore significant not because each device is fitted with an independent accommodation coefficient, but because the dry-out heat fluxes of different devices are reproduced using a single fixed value of $\sigma_{\mathrm{ac}}=0.054$ together with contact angles constrained by the interferometric measurements of Adera et al.~\cite{Adera.2016}.

For water, reported accommodation coefficients span a wide range, from values of order $10^{-2}$ to values approaching unity, depending on the measurement method and modeling assumptions \cite{Marek.2001}. One likely reason is that the microscopic liquid-vapor interface shape, which strongly affects the local thermal resistance and evaporative flux distribution, is not always resolved explicitly. Within this context, the ability of the present model to reproduce the measured dry-out heat fluxes of different devices using a single fixed accommodation coefficient and contact angles within the experimentally observed range supports the internal consistency of the thermal model.

\section{Appendix: Validation of dry-out calculations against Zhu et al.}
\label{app:Zhu16_dry_out_validation}
\begin{figure}[pos=htbp]
\centering
\includegraphics{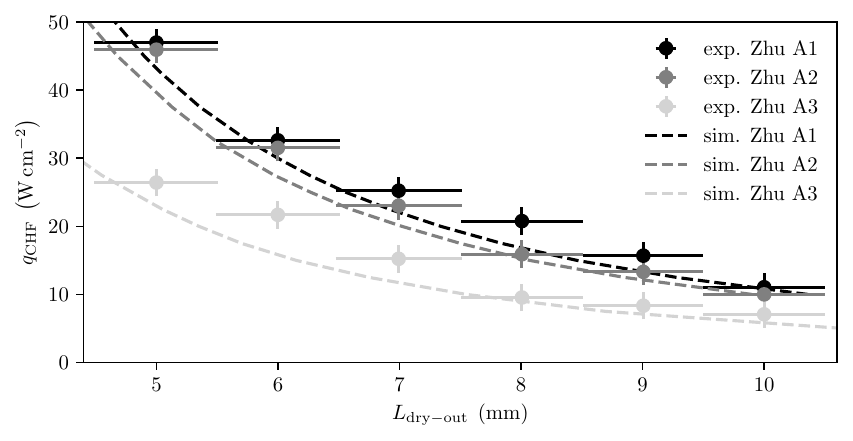}
\caption{Comparison between experimental and simulated dry-out data for the water-filled silicon micropillar wicks with thermal oxide coating reported by Zhu et al.~\cite{Zhu.2016}. The experiments were performed at $p=\qty{101.3}{\kilo\pascal}$ and $T\approx\qty{373.2}{\kelvin}$. The comparison includes devices A1 ($d=\qty{7}{\micro\meter}$, $l=\qty{20}{\micro\meter}$, $h=\qty{30}{\micro\meter}$), A2 ($d=\qty{7}{\micro\meter}$, $l=\qty{30}{\micro\meter}$, $h=\qty{19}{\micro\meter}$), and A3 ($d=\qty{6}{\micro\meter}$, $l=\qty{50}{\micro\meter}$, $h=\qty{19}{\micro\meter}$). The model shows near-quantitative agreement with the experimental dry-out data within the reported uncertainty. Temperature-dependent water surface-tension data were taken from Ref.~\cite{Vargaftik.1983}, a receding contact angle of $\theta_{\mathrm{r}}\approx \qty{10}{\degree}$ was assumed following Ref.~\cite{Raj.2013}, and the temperature-dependent latent heat of vaporization used in the thermal simulations was taken from Ref.~\cite{CRCHandbook.2004}. The remaining thermophysical properties were obtained from the COMSOL water material-property database.}
\label{fig:Zhu16_comparison}
\end{figure}

To validate the dry-out calculation against an established experimental reference, the model was applied to the water-filled silicon micropillar wicks with thermal oxide coating reported by Zhu et al.~\cite{Zhu.2016}. As shown in Fig.~\ref{fig:Zhu16_comparison}, the parameter-free model reproduces the measured dry-out data for the three reported devices with near-quantitative agreement, remaining within the experimental uncertainty.

\section{Derivation of the Approximate Dry-Out Model}
\label{app:dryout_scaling_derivation}

This appendix presents the derivation of the approximate capillary-viscous dry-out model used in Eq.~\eqref{eq:dryout_scaling_estimate}. Similar more qualitative scaling law formulations for dry-out in liquid-fed micropillar wicks have been proposed previously (e.g., Zhu et al.~\cite{Zhu.2016}). The present derivation serves as an order-of-magnitude model and neglects spatial variations in the temperature field and meniscus shape.

We consider a single-fed square micropillar array assumed to be periodic in the transverse $y$-direction. We analyze a single unit-cell-wide stripe extending along the $x$-direction from the reservoir at $x=0$ to the dry-out front at $x = L_{\mathrm{dry\text{-}out}} = Nl$, where $N$ is the total number of cells and $l$ is the pillar pitch.

The liquid mass flow rate through cell $n \in \{1, \dots, N\}$ must supply the cumulative evaporation from cell $n$ through cell $N$. For a uniform heat flux $q_{\mathrm{heat}}$, the mass flow rate is given by
\begin{equation}
    \label{eq:app_dryout_mass_flow_cell}
    \dot m_n = \frac{q_{\mathrm{heat}} l^2 \left(N - n + 1\right)}{h_{\mathrm{lv}}} \;,
\end{equation}
and the corresponding volumetric flow rate is $\dot V_n = \dot m_n / \rho_{\mathrm{l}}$.

Assuming a Darcy-type pressure relation and taking a constant cell-geometry-dependent permeability $k_{d,l,h}$ across the wick, the pressure drop across cell $n$ with cross-sectional flow area $lh$ is
\begin{align}
    \label{eq:app_dryout_cell_pressure_drop}
    \Delta p_n &= \frac{\mu_{\mathrm{l}}}{k_{d,l,h}} \frac{l}{lh} \dot V_n \nonumber \\
    &= \frac{\mu_{\mathrm{l}}}{h k_{d,l,h}} \frac{q_{\mathrm{heat}} l^2 \left(N - n + 1\right)}{\rho_{\mathrm{l}} h_{\mathrm{lv}}} \;.
\end{align}
Summing over all $N$ cells yields the total viscous pressure drop along the wick:
\begin{align}
    \label{eq:app_dryout_total_pressure_drop_exact}
    \Delta p &= \sum_{n=1}^{N} \Delta p_n \nonumber \\
    &= \frac{\mu_{\mathrm{l}} q_{\mathrm{heat}} l^2}{\rho_{\mathrm{l}} h_{\mathrm{lv}} h k_{d,l,h}} \frac{N(N+1)}{2} \;.
\end{align}
For sufficiently large $N$ ($N \gg 1$), this simplifies to
\begin{equation}
    \label{eq:app_dryout_total_pressure_drop_large_N}
    \Delta p \approx \frac{\mu_{\mathrm{l}} q_{\mathrm{heat}} l^2 N^2}{2 \rho_{\mathrm{l}} h_{\mathrm{lv}} h k_{d,l,h}} \;.
\end{equation}
Dry-out is estimated by setting this viscous pressure drop equal to the critical capillary pressure ($\Delta p = P_{\mathrm{crit}}$). Substituting Eq.~\eqref{eq:Critical_pressure_sparse_array} for the critical capillary pressure of a sparse pillar array yields
\begin{equation}
    \label{eq:app_dryout_length_squared}
    N^2 l^2 \approx 8 \frac{ \sigma_{\mathrm{lv}} \rho_{\mathrm{l}} h_{\mathrm{lv}}}{\mu_{\mathrm{l}} q_{\mathrm{heat}}} \frac{\dfrac{h}{d} k_{d,l,h}}{\dfrac{4}{\pi} \left( \dfrac{l}{d} \right)^2 - 1} \cos(\theta_{\mathrm{r}}) \;.
\end{equation}
Using $L_{\mathrm{dry\text{-}out}} = Nl$ and taking the square root yields Eq.~\eqref{eq:dryout_scaling_estimate}.







\clearpage 

\printcredits

\bibliographystyle{elsarticle-num} 
\bibliography{cas-refs}



\end{document}